\documentclass{article}
\usepackage{graphicx}
\usepackage{amsmath}
\usepackage{verbatim}
\usepackage{url}
\usepackage{geometry}
\usepackage{lscape}
\usepackage{subfigure}
\usepackage[authoryear, round, sort]{natbib}
\bibpunct{(}{)}{;}{a}{}{,}

\begin{document}
\title{Quantifying Selection and Diversity in Viruses by Entropy Methods, with Application to the Hemagglutinin of H3N2 Influenza}
\author{$\mbox{Keyao Pan}^1$ and $\mbox{Michael W. Deem}^{1,2}$\\\\
Department of $^1$Bioengineering and $^2$Physics \& Astronomy, Rice University,\\
6100 Main Street, Houston, TX 77005}

\date{}

\maketitle

\begin{abstract}
Many viruses evolve rapidly. For example, hemagglutinin of the H3N2 influenza A virus evolves to escape antibody binding. This evolution of the H3N2 virus means that people who have previously been exposed to an influenza strain may be infected by a newly emerged virus. In this paper, we use Shannon entropy and relative entropy to measure the diversity and selection pressure by antibody in each amino acid site of H3 hemagglutinin between the 1992--1993 season and the 2009--2010 season. Shannon entropy and relative entropy are two independent state variables that we use to characterize H3N2 evolution. The entropy method estimates future H3N2 evolution and migration using currently available H3 hemagglutinin sequences. First, we show that the rate of evolution increases with the virus diversity in the current season. The Shannon entropy of the sequence in the current season predicts relative entropy between sequences in the current season and those in the next season. Second, a global migration pattern of H3N2 is assembled by comparing the relative entropy flows of sequences sampled in China, Japan, the USA, and Europe.  We verify this entropy method by describing two aspects of historical H3N2 evolution. First, we identify 54 amino acid sites in hemagglutinin that have evolved in the past to evade the immune system. Second, the entropy method shows that epitopes A and B in the top of hemagglutinin evolve most vigorously to escape antibody binding. Our work provides a novel entropy-based method to predict and quantify future H3N2 evolution and to describe the evolutionary history of H3N2.
\end{abstract}


\newpage

\section{Introduction}
\label{sec:Introduction}

\begin{figure}
\centering
\includegraphics[width=2in]{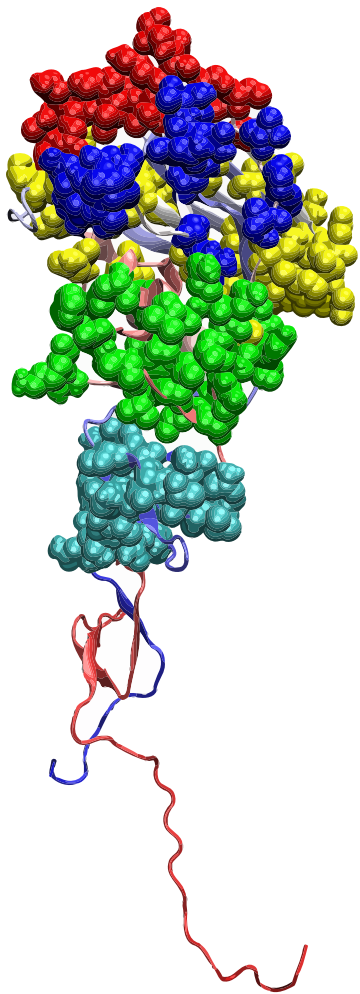}
\caption{The tertiary structure of the HA1 domain of H3 hemagglutinin (PDB code: 1HGF). The surface of HA1 facing outward is the exposed surface when the hemagglutinin trimer is formed. The other two HA1 domains (not shown) in the HA trimer are located at the back of the structure displayed here. The solid balls represent five epitopes. Color code: blue is epitope A, red is epitope B, cyan is epitope C, yellow is epitope D, and green is epitope E.}
\label{fig:H3_epitope}
\end{figure}

A common strategy by which viruses evade pressure from the immune system is to evolve and change their antigenic profile.  Viruses with a low evolutionary rate that infect only humans, such as the small pox virus \citep{Li2007}, can be effectively controlled by vaccinating the human population.  By contrast, viruses with a high evolutionary rate, such as HIV, hepatitis B, and influenza A, resist being eliminated by the immune system by generating a plethora of mutated virus particles and causing chronic or repeated infection.  In this study, we take subtype H3N2 influenza A virus as a model evolving virus. Influenza A virus circulates in the human population every year, typically causing 3--5 million severe illnesses and 250,000--500,000 fatalities all over the world \citep{WHO2009}. Hemagglutinin (HA) and neuraminidase (NA) are two kinds of virus surface glycoproteins encoded by the influenza genome. The subtype of influenza is jointly determined by the type of hemagglutinin ranging from H1 to H16, and that of neuraminidase ranging from N1 to N9. On the surface of the virus membrane, HA exists as a cylindrical trimer containing three HA monomers, and each monomer comprises two domains, HA1 and HA2. Hemagglutinin is also a key factor in virus evolution, because it is the major target of antibodies, and HA escape mutation changes the antigenic character of the virus presented to the immune system. The H3N2 virus causes the largest fraction of influenza illness. H3 hemagglutinin is under selection by the immune response mainly on the five epitope regions in the HA1 domain \citep{Wiley1981}, labeled epitopes A to E, as shown in Figure \ref{fig:H3_epitope}. The immune pressure and the escape mutation drive the evolution of the H3N2 virus. The underlying mutation rate of the HA gene is $1.6 \times 10^{-5}/\mbox{amino acid position}/\mbox{day}$ \citep{Nobusawa2006}, measured using the method modified from that in an earlier study on the HA mutation rate \citep{Parvin1986}. Note that the mutation rate does not necessarily equal to the evolution rate, or the fixation rate. The mutation rate equals to the evolutionary rate only if the evolution is neutral. The non-neutrality of the HA evolution is shown in the Results section. Evolution of the hemagglutinin viral protein causes occasional mismatch between the virus and the vaccine and decreases vaccine effectiveness \citep{Gupta2006,Deem2003}. As more amino acid substitutions are introduced into influenza sequences, the antigenic characteristics of influenza strains drift away \citep{Liao2008}, and influenza epidemic severity of subtype H1N1 \citep{Wu2010} and subtype H3N2 \citep{Wolf2010} increases.

The H3N2 virus has a distinguished evolutionary history, largely affected by the immune pressure. The H3N2 virus emerged in the human population in 1968 and has been circulating in the population since 1968. The phylogenetic tree of H3 hemagglutinin since 1968 has a linear topology in which most sequences are close to the single trunk of the tree, and the lengths of the branches are short \citep{Ferguson2003,Smith2004,Russell2008}. Historical hemagglutinin sequences fall into a series of clusters, each of which has similar genetic or antigenic features and circulates for 2--8 years before being replaced by the next cluster \citep{Plotkin2002,Smith2004}. The evolution of different amino acid positions of hemagglutinin shows a remarkable heterogeneity: a subset of positions undergo frequent change, while some positions are conserved \citep{Shih2007}. This heterogeneity is quantified by the Shannon entropy at each position of the amino acid sequence of hemagglutinin \citep{Deem2009}. Shannon entropy has been used to locate protein regions with high diversity, such as the antigen binding sites of T-cell receptors \citep{Stewart1997}. Shannon entropy has been used to identify antibody binding sites, or epitopes, which are under immune pressure and so are rapidly evolving \citep{Deem2009}. The heterogeneity of amino acid substitution suggests that point mutations randomly occurring in distinct positions have different contributions to the virus fitness.


The selection pressure on the H3N2 virus to evolve is reflected in the difference between the H3 hemagglutinin sequences in two consecutive seasons. We consider Northern hemisphere strains. When the epidemic initiates in a new season, we assume that each position of an HA sequence inherits the amino acid from a sequence of the previous season or has a different amino acid due to random mutation and selection. This assumption comes from the fact that the H3N2 virus circulating in each influenza season migrates from a certain geographic region in which the virus is preserved between two influenza seasons \citep{Russell2008,Rambaut2008}. In the absence of selection, the histogram of the 20 amino acid usage in one position in the current season is similar to that in the same position in the previous season except for changes due to the small random mutation rate. The difference between the two histograms beyond that expected due to mutation quantifies selection.

Synthesizing these factors, we introduce an entropy method to describe the evolution of influenza.  The entropy method extracts an evolutionary pattern from aligned sequences. Shannon entropy quantifies the amount of sequence information in each position of aligned sequences \citep{Schneider1986,Schneider1990}. The sequence information reflects the variation, which is equivalently diversity, in each position, and so Shannon entropy has been used to measure the diversity in each position \citep{Sander1991,Shenkin1991,Gerstein1995}. Shannon entropy has also been used to measure the structural conservation in the protein folding dynamics \citep{Mirny1999,Plaxco2000}. See \citep{Valdar2002} for a detailed review of the applications of Shannon entropy. On the other hand, relative entropy measures gain of sequence information at each position and requires a background amino acid frequency distribution \citep{Williamson1995}. Relative entropy was also used as a sequence conservation measure to detect functional protein sites \citep{Wang2006,Halabi2009}. Further, a dimension reduction technique using relative entropy has identified sectors in proteins \citep{Lockless1999,Halabi2009}. As an extension of these previous works, we apply Shannon entropy and relative entropy to jointly measure two quantities in each position: sequence information in one season and gain of sequence information from one season to the next season. Simultaneous analysis of Shannon entropy and relative entropy sheds light on the evolutionary pattern of the H3N2 virus evolution when data from multiple seasons are available.  In the HA1 domain, positions in the epitope regions have increased Shannon entropy, and this feature was applied to locate the epitopes of H1 hemagglutinin \citep{Deem2009}.  We here use Shannon entropy to quantify the virus diversity in each amino acid position in each season.  The entropy relative to the previous season \citep{Kullback1951} is also used to analyze the evolution of the HA1 domain in one single season and to quantify the selection pressure on the virus in each amino acid position in each season.  The selection and the virus diversity are two significant state variables determining the dynamics of evolution.




The article is organized as follows. The Materials and Methods section presents the data used in this work and details of the entropy model. The Results section uses Shannon entropy of the sequence in one season to predict the evolution quantified by relative entropy from this season to the next season. Results are also presented for the flow of virus migration between the four geographic regions of China, Japan, the USA, and Europe. In the Result section, we demonstrate the entropy method in two applications, the results of which agree with prior knowledge on H3N2 evolution. Finally, we discuss our results and present our conclusions.

\section{Materials and Methods}
\label{sec:Materials_and_Methods}

\subsection{Sequence Data}
\label{sec:Sequence_Data}

The hemagglutinin sequences considered in this work are the amino acid sequences of the hemagglutinin of human influenza A H3N2 virus. We only use Northern Hemisphere sequences because 90\% of the world population lives in Northern Hemisphere. The influenza season in Northern Hemisphere is defined as the time interval from October in one year to September in the next year. We downloaded 5309 Northern Hemisphere sequences labeled with month of collection from the NCBI Influenza Virus Resource on 16 January 2011. Sequences too short to cover residues 1--329 of hemagglutinin were removed, and the remaining sequences were trimmed to only keep residues 1--329 in the HA1 domain. Any sequence with an undefined amino acid in residues 1--329 was removed. We consider 18 seasons from 1992--1993 to 2009--2010 during which most available sequences were collected. In total, we preserved and aligned 4292 sequences in these 18 seasons containing amino acids 1--329.

\subsection{Histograms of 20 Amino Acids}
\label{sec:Histograms_of_20_Amino_Acids}

The first step is to quantify the alignment of the amino acid sequences. The aligned historical H3 hemagglutinin sequences form a matrix $\mathbf{A}$ with 4292 rows and 329 columns. The element $\mathbf{A}_{l,j}$ denotes the identity of the amino acid in sequence $l$ and position $j$. The 4292 sequences were clustered into 18 groups by the seasons of sampling from the 1992--1993 season to the 2009--2010 season. Note that most of the sequences before the 1992--1993 season were not labeled with month of collection and are excluded from this study. We denote by $i=0,1,\dots,17$ the seasons between 1992--1993 and 2009--2010. For position $j$ in season $i$, the relative frequency of each amino acid $k$, $f\left(k,i,j\right), k=1,\dots,20$, was calculated from the vector $\mathbf{A}_{\vec{l}\left(i\right),j}$ where the index array $\vec{l}\left(i\right)$ holds the indices of sequences sampled in season $i$.

\subsection{Shannon Entropy as Diversity}
\label{sec:Diversity}

The Shannon entropy is one useful quantification of the diversity in single position. Large Shannon entropy has the physical meaning that the amino acid in the given position is prone to be substituted. This physical meaning was also applied in \citep{Deem2009}. The diversity at a single position takes the format of Shannon entropy because of the sensitivity of Shannon entropy to diversity.

This physical meaning of the Shannon entropy does not necessarily involve the joint frequency distributions for two and more positions, and we do not consider the joint frequency in the manuscript. Rather, we define the diversity only in the level of single amino acid position. Consequently, the defined diversity is additive for a number of positions. The idea of adding diversity in each position of the sequence comes from classic works such as \citep{Schneider1986}, which added the Shannon entropy in each position to measure the total diversity in an aligned binding site.

Therefore, diversity of the virus in each position in each season is represented by the Shannon entropy that quantifies the amount of information in the histogram or distribution under study. For the sequences sampled in all the seasons, positions with high evolutionary rate have a higher Shannon entropy compared to the conserved positions \citep{Deem2009}. The sequences in each season are assumed to be collected concurrently. The Shannon entropy is a quantification of diversity of amino acids in one position, and so the diversity in position $j$ in season $i$ is calculated from the histogram $\mathbf{f}\left(i,j\right) = \left[f\left(1,i,j\right), \dots, f\left(20,i,j\right)\right]^\mathrm{T}$ by Shannon entropy
\begin{equation}\label{eq:diversity}
D_{i,j} = -\sum_{k=1}^{20} f\left(k,i,j\right) \log f\left(k,i,j\right)
\end{equation}
in which $k = 1, \dots, 20$ is the identity of the amino acid in position $j$ in season $i$.

\subsection{Relative Entropy as Selection Pressure}
\label{sec:Selection}

Selection in each position $j$ in season $i$ is reflected by the significant difference between the 20-bin histogram in the current season $\mathbf{f}\left(i,j\right)$ and that in the previous season $\mathbf{f}\left(i-1,j\right)$. In the absence of selection, random mutation and genetic drift are the dominant forces generating $\mathbf{f}\left(i,j\right)$ from $\mathbf{f}\left(i-1,j\right)$. In each position, random mutation creates a slightly modified histogram, from which amino acids are randomly chosen to appear in season $i$ by the effect of genetic drift.

The source of random mutation is the spontaneous error of the RNA polymerase replicating the influenza virus RNA. The random mutation rate in different regions of hemagglutinin is thought to be homogeneous, regardless whether the regions are in antigenic sites or not \citep{Ina1994}. Therefore random mutation is modeled as a Poisson process $\mathbf{M}\left(\mu, g\left(k\right)\right)$ equally affecting all the positions. Here $\mu$ is the mutation rate of influenza A virus that equals to $5.8 \times 10^{-3}/\mathrm{residue}/\mathrm{season}$ \citep{Nobusawa2006}, and $g\left(k\right), k=1,\dots,20$ is the relative frequency of each amino acid in the whole alignment $\mathrm{A}$. The probability that the original amino acid $t$ mutates to amino acid $u$ is
\begin{equation}\label{eq:mutation}
\mathbf{M}_{u,t} \left(\mu, g\right) = \frac{\mu g\left(u\right)}{1 - g\left(t\right)}.
\end{equation}
So after mutating for one season, the histogram in position $j$ in season $i-1$ is obtained by
\begin{equation}\label{eq:background}
\hat{\mathbf{f}}\left(i,j\right) = \mathbf{M} \left(\mu, g\right) \mathbf{f}\left(i-1,j\right).
\end{equation}
This histogram serves as the background distribution for season $i$ from which the sequences in season $i$ are built.

If selection is absent, the effect of genetic drift is to create sequences in the current season by randomly choosing amino acids in each position from a background distribution $\hat{\mathbf{f}}\left(i,j\right)$. We denote by $N_i$ the number of sequence in season $i$. The probability that $N_i$ amino acids in position $j$ have the histogram $\mathbf{f}\left(i,j\right)$ is \citep{Halabi2009}
\begin{equation}\label{eq:sample_hist}
\mathrm{Pr}\{\mathbf{f}\left(i,j\right)\} \approx \exp\left(-N_i S_{i,j}\right)
\end{equation}
where
\begin{equation}\label{eq:selection}
S_{i,j} = \sum_{k=1}^{20} f\left(k,i,j\right) \log\frac{f\left(k,i,j\right)}{\hat{f}\left(k,i,j\right)}
\end{equation}
is the relative entropy between the observed histogram, $f\left(k,i,j\right)$, and the background histogram, $\hat{f}\left(k,i,j\right)$ \citep{Kullback1951}.

The null hypothesis that selection is absent in the evolution is rejected if the relative entropy $S_{i,j}$ is great enough such that the probability in Equation \ref{eq:sample_hist} is less than 0.05, that is, the relative entropy $S_{i,j}$ is greater than $-\log\left(0.05\right) / N_i$ in season $i$. Note that the majority of residues were stable in most of the seasons, and in this case the relative entropy is $S_{i,j} = \log\left(1/\left(1-\mu\right)\right) \approx \mu$. To avoid classifying these stable residues erroneously as positions under selection, a proper threshold of relative entropy needs to be larger than the mutation rate $\mu$. Additionally, a fraction $\lambda$ of the circulating HA1 sequences were not deposited in the database because of the sampling bias of the HA1 sequences. In an extreme case, in a stable position $j$ with the real histogram of 20 amino acids $\left[1-\lambda,\lambda,\dots,0\right]^\mathrm{T}$ in all the seasons, the histograms of the sequences sampled in two consecutive seasons $i-1$ and $i$ are $\mathbf{f}\left(i-1,j\right) = \left[1,0,\dots,0\right]^\mathrm{T}$ and $\mathbf{f}\left(i,j\right) = \left[1-\lambda,\lambda,\dots,0\right]^\mathrm{T}$, respectively, and so the relative entropy introduced by the sampling bias is
\begin{equation}
S^\mathrm{bias} \approx \left(1-\lambda\right) \log\frac{1-\lambda}{1-\mu} + \lambda \log\frac{\lambda}{\mu/19}
\end{equation}
in spite of the absence of selection in position $j$ in season $i$.
The relative entropy $S^\mathrm{bias}$ equals to 0.1 if a sampling bias $\lambda = 2.5\%$ exists in the HA1 database sequences.
We fix the threshold of the relative entropy in season $i$ to
\begin{equation}\label{eq:threshold}
S_i^\mathrm{thres} = \max\left\{-\log\left(0.05\right) / N_i,\left(1-\lambda\right) \log\frac{1-\lambda}{1-\mu} + \lambda \log\frac{\lambda}{\mu/19}\right\} \approx \max\left\{3/N_i, 0.1\right\}.
\end{equation}
The numbers of collected HA1 sequences $N_i$ were fewer than 30 only in the 1995--1996 season $\left(i=3\right)$ with $N_3 = 25$. The thresholds $S_3^\mathrm{thres} = 0.12 > 0.1$ in the 1995--1996 season due to the small numbers of HA1 sequences. In all the other 17 seasons, the numbers of sequences $N_i$ were greater than 30, and so the thresholds $S_i^\mathrm{thres} = 0.1$.

\section{Results}
\label{sec:Results}

In this section, we show the positive correlation between the Shannon entropy in season $i$ and the relative entropy from season $i$ to season $i+1$. This correlation means that the larger the virus diversity in one season, the higher the virus evolutionary rate from this season to the next season. We draw the H3N2 migration pattern by comparing the relative entropy. The migration pattern reveals a novel migration path from the USA to Europe and shows that the virus evolutionary rate is higher in the epicenter, China, than in the migration paths. We also demonstrate the entropy method in two applications. First, we compute average Shannon entropy and relative entropy in each position over the past 17 seasons to identify positions under selection pressure. Second, we compare Shannon entropy and relative entropy in epitope regions to find the contribution of each epitope to the H3N2 evolution. Results of these two applications agree with previous studies and additionally show the heterogeneity of the selection pressure over different amino acid positions of hemagglutinin, with increased pressure in the epitopes, as well as the dominance of epitopes A and B.

\subsection{Correlation between Shannon entropy and relative entropy}
\label{sec:Correlation_between_entropy}

\begin{figure}
\centering
\includegraphics[width=5in]{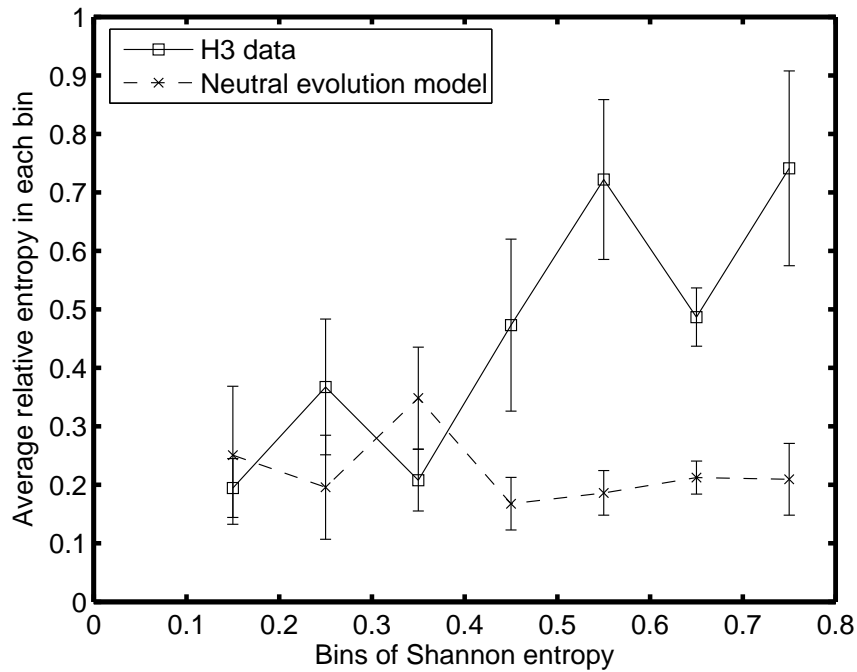}
\caption{Mean and standard error of relative entropy $S_{i+1,j}$ in each bin of Shannon entropy. Shannon entropy and relative entropy in each of the 329 positions and in each of the 17 seasons between 1992--1993 ($i = 0$) and 2008--2009 ($i = 16$) fall into one of the eight bins. The first bin with Shannon entropy less than 0.1 is discarded.  Bins with larger Shannon entropy $D_{i,j}$ also have larger relative entropy $S_{i+1,j}$.  Shannon entropy $D_{i,j}$ and relative entropy $S_{i+1,j}$ in iteration $i =$ 51--100 of the neutral evolution model are used to calculate mean and standard error of relative entropy in each bin of Shannon entropy distribution in the same way. No increasing trend is found. Error bar is one standard error.}
\label{fig:entropy_allposition}
\end{figure}

Relative entropy $S_{i+1,j}$ in amino acid position $j$ from season $i$ to season $i+1$ linearly increases with Shannon entropy $D_{i,j}$ in position $j$ in season $i$. For the sequences sampled from the 1992--1993 season ($i = 0$) to the 2008--2009 season ($i = 16$), $329 \times 17 = 5593$ ordered pairs $(D_{i,j}, S_{i+1,j})$ are calculated. All except two pairs fall into 8 bins in which the values of $D_{i,j}$ belong to 8 intervals $[0, 0.1)$, $[0.1, 0.2)$, $\dots$, $[0.7, 0.8)$, respectively. The first bin with $D_{i,j}$ in $[0, 0.1)$ is discarded because it contains numerous conserved amino acid positions. The values of $S_{i+1,j}$ are averaged respectively in each of the 7 remaining bins. As described in Figure \ref{fig:entropy_allposition}, average relative entropy in each bin shows positive correlation with midpoints of the $D_{i,j}$ interval, $R^2 = 0.70$. An amino acid position $j$ with high Shannon entropy $D_{i,j}$ in season $i$ is expected to present high relative entropy $S_{i+1,j}$ from season $i$ to $i+1$. The evolution in position $j$ from season $i$ to $i+1$ quantified by relative entropy is therefore predicted using the mean and standard error of $S_{i+1,j}$ in the bin chosen by $D_{i,j}$ in season $i$.

\begin{figure}
\centering
\includegraphics[width=5in]{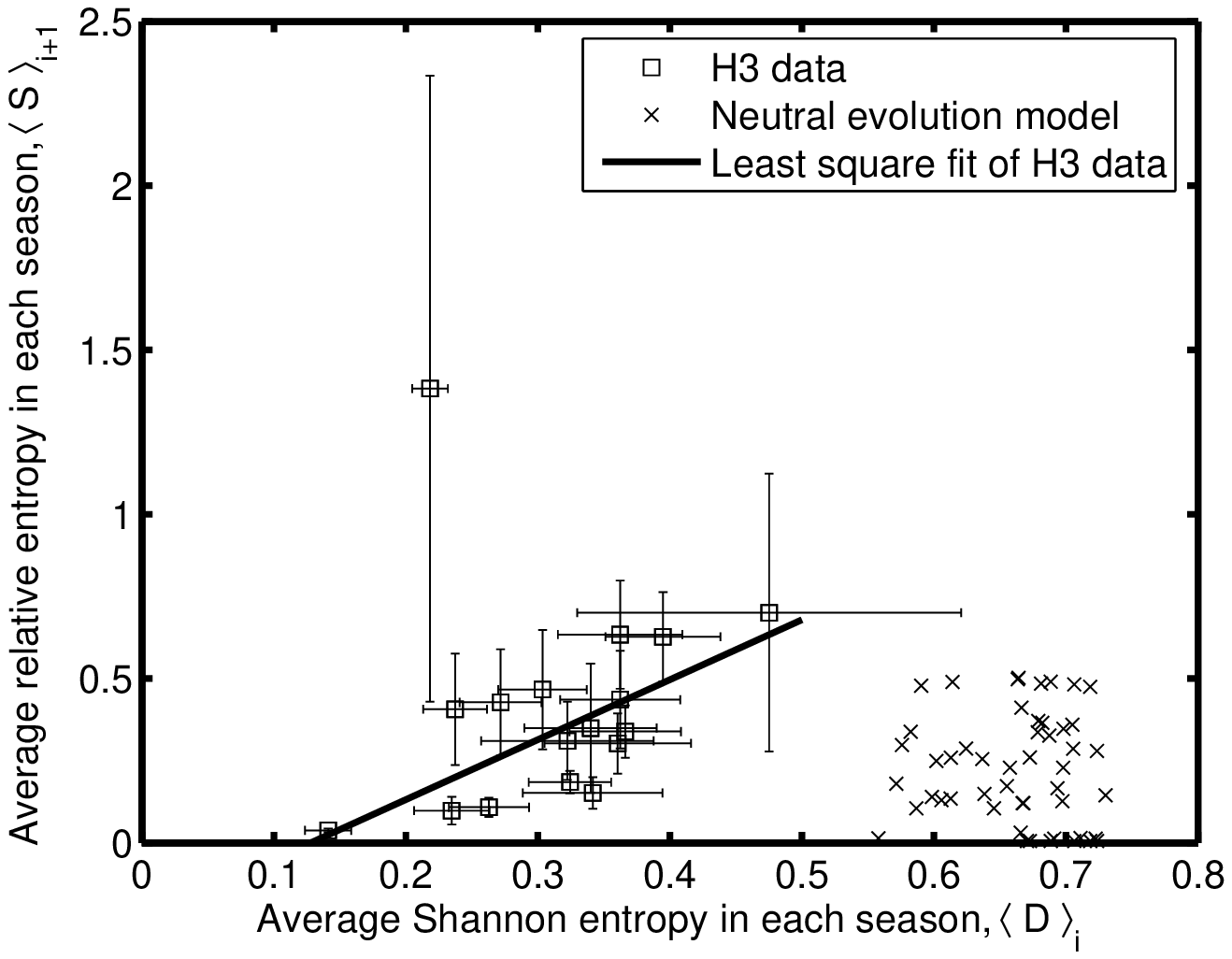}
\caption{Average Shannon entropy $\langle D \rangle_i$ versus average relative entropy $\langle S \rangle_{i+1}$ for each season between 1992--1993 ($i = 0$) and 2008--2009 ($i = 16$).  For each season $i$, a set of amino acid positions $j$ with Shannon entropy $D_{i,j}$ greater than 0.1 are chosen. For all the $j$ in this set of positions, $\langle D \rangle_i$ is the average of the Shannon entropy $D_{i,j}$ values and $\langle S \rangle_{i+1}$ is the average of relative entropy $S_{i+1,j}$ values. Horizontal and vertical error bars are the standard errors of Shannon entropy and relative entropy, respectively. The solid line, $\langle S \rangle_{i+1} = 1.82 \langle D \rangle_{i} - 0.23$, is a least squares fit of $\langle D \rangle_i$ to $\langle S \rangle_{i+1}$ ($i = 0, 2, \dots, 16$). A strong correlation with $R^2 = 0.50$ exists between $\langle D \rangle_i$ and $\langle S \rangle_{i+1}$ excluding the point $\left(0.22, 1.38\right)$ with $N_i = 1$, which has a large standard error of the relative entropy $S_{i+1,j}$. Using the same method, $\langle D \rangle_i$ and $\langle S \rangle_{i+1}$ are calculated from a neutral evolution model, $i = $ 51--100, and plotted.  No visible correlation exists between $\langle D \rangle_i$ and $\langle S \rangle_{i+1}$ from the neutral evolution model.}
\label{fig:entropy_avg_year}
\end{figure}

Positive correlation is also observed between the mean values of $D_{i,j}$ and $S_{i+1,j}$ in a variety of positions $j$ in each season $i$. In each season between 1992--1993 ($i = 0$) and 2008--2009 ($i = 16$), we average the Shannon entropy $D_{i,j}$ and relative entropy $S_{i+1,j}$ over the positions $j$ with $D_{i,j} > 0.1$. The data point with $i=1$, $\left(0.22, 1.38\right)$, has a large standard error of the relative entropy $S_{i+1,j}$ and is excluded in the analysis below. The remaining average Shannon entropy $\langle D \rangle_{i}$ ($i = 0, 2, \dots, 16$) correlates with average relative entropy $\langle S \rangle_{i+1}$ ($i = 0, 2, \dots, 16$) with $R^2 = 0.50$, as shown in Figure \ref{fig:entropy_avg_year}. A least squares fit gives $\langle S \rangle_{i+1} = 1.82 \langle D \rangle_{i} - 0.23$. Thus, the expected average relative entropy from the current season $i$ to the next season $i+1$ can be calculated from the average Shannon entropy $\langle D \rangle_{i}$ in the current season $i$.

The above relationships between Shannon entropy and relative entropy cannot be generated by a neutral evolution model. To demonstrate this result, we create an ensemble of 1000 identical sequences with 50 amino acid positions. Each iteration of the model simulates H3N2 evolution during one season. In each iteration, the number of mutated amino acids $N_\mathrm{mut}$ in each sequence follows a Poisson distribution with mean $\mu = 2.0$, which is the annual substitution rate in history. The $N_\mathrm{mut}$ mutated positions are then randomly assigned in the corresponding sequence. We randomly select $p_\mathrm{cut} = 10\%$ of the sequences to build the sequence ensemble in the next iteration. The Shannon entropy $D_{i,j}$ and relative entropy $S_{i+1,j}$ generated in iteration $i = $ 51--100 are processed using the same method as for H3 sequences in history. First, as shown in Figure \ref{fig:entropy_allposition}, no increasing trend appears in the means of $S_{i+1,j}$ in the 7 bins from $[0.1, 0.2)$ to $[0.7, 0.8)$. Second, no correlation ($R^2 = 0.003$) is observed between $\langle D \rangle_{i}$ and $\langle S \rangle_{i+1}$, see Figure \ref{fig:entropy_avg_year}. When we change the parameters $\mu$ between 1.0 and 10 and $p_\mathrm{cut}$ between 1\% and 100\% in the algorithm, the simulation still does not yield the visible increasing trend in Figure \ref{fig:entropy_allposition} or the correlation observed in Figure \ref{fig:entropy_avg_year}. As a result, we conclude that neutral evolution alone is not able to generate the pattern between Shannon entropy and relation entropy of H3 sequences. It was previously shown that the fixation rate of H3N2 evolution cannot be explained only by neutral evolution \citep{Shih2007}. In this study, the monotonically increasing linear relationship between relative entropy and Shannon entropy in Figure \ref{fig:entropy_allposition} and \ref{fig:entropy_avg_year} suggests that selection pressure substantially contributes to H3N2 evolution.

\subsection{Annual Virus Migration}
\label{sec:Prediction_of_Future_Virus_Strains}

The entropy method is also used to analyze the global migration pattern of the virus.  Most of the Northern Hemisphere H3 sequences were collected in East-Southeast Asia, the USA, and Europe. East-Southeast Asia is suggested to be the reservoir of the annual H3N2 epidemic \citep{Russell2008}. To increase the geographic resolution in East-Southeast Asia, we use two regions, China and Japan, as the representative of East-Southeast Asia, because each of these regions has a population over 50 million and has a consistent time series of H3 sequence data from the 2001--2002 to the 2007--2008 season.

We select the sequences in four geographic regions that are China, Japan, the USA, and Europe in seven seasons from 2001--2002 to 2007--2008.  In all the six pairs of consecutive seasons, we calculate for each region four average relative entropy values of the whole sequence. These four average relative entropy values are calculated using the sequences in each of the four regions in the previous season as the reference.  The results are shown in Table \ref{tab:migration}.  Sequences collected in China in the previous season yield the minimum relative entropy to the sequences in the current season collected in China ($p < 2.1 \times 10^{-5}$, Wilcoxon signed-rank test), Japan ($p < 0.0049$, Wilcoxon signed-rank test), and the USA ($p < 0.0012$, Wilcoxon signed-rank test).  Sequences in the USA in the previous season have the minimum relative entropy to the sequences in Europe in the current season ($p < 0.15$, Wilcoxon signed-rank test).  Relative entropy data in Table \ref{tab:migration} imply the virus migration from China, as the geographic reservoir, to Japan and the USA and suggest a migration from the USA to Europe.  The result in Table \ref{tab:migration} also implies that the H3N2 virus circulating in China seed the virus in China, Japan, and the USA in the next season, and the virus in the USA probably seed the virus in Europe in the next season.

\begin{table}
\caption{The relative entropy between hemagglutinin sequences in the different regions in the current influenza season and sequences in these regions in the previous season.  The minimum relative entropy in each column is marked in bold.  The $p$ values of the Wilcoxon signed-rank test between the minimum relative entropy and other relative entropy values in the same column are in the parentheses. Hemagglutinin sequences were collected from four geographic regions: China, Japan, the USA, and Europe.  Seven seasons from 2001--2002 to 2007--2008 are used here.  The relative entropy values listed in this table are averaged for all the sites and all the six pairs of consecutive seasons. These results imply that the H3N2 viruses in China, Japan, and the USA migrate from China, while the H3N2 virus in Europe migrates from USA.} \centering
\begin{tabular}{l l l l l l}
\\\hline
                & Region of the   &        &        &         &        \\
                & current season  & China  & Japan  & USA     & Europe \\
Region of the   &                 &        &        &         &        \\
previous season &                 &        &        &         &        \\\hline
China           &                 & \textbf{0.057}  & \textbf{0.040}  & \textbf{0.044}   & 0.064  \\
                &                 &                 &                 &                  & (0.0017) \\\hline
Japan           &                 & 0.114  & 0.094  & 0.076   & 0.059  \\
                &                 & ($2.1 \times 10^{-5}$) & (0.0049) & (0.0012) & (0.032) \\\hline
USA             &                 & 0.105  & 0.087  & 0.070   & \textbf{0.056}  \\
                &                 & ($2.1 \times 10^{-10}$) & ($3.3 \times 10^{-8}$) & ($6.4 \times 10^{-5}$) & \\\hline
Europe          &                 & 0.135  & 0.115  & 0.094   & 0.074  \\
                &                 & ($3.8 \times 10^{-9}$) & ($3.4 \times 10^{-6}$) & ($4.8 \times 10^{-7}$) & (0.15) \\
\hline
\end{tabular}
\label{tab:migration}
\end{table}

Comparison of the relative entropy data in Table \ref{tab:migration} in the H3N2 reservoir and migration paths also reveals the H3N2 virus migration pattern. Using the Wilcoxon sign-rank test, the relative entropy data of the H3 hemagglutinin in China in two consecutive seasons is significantly greater than those in the three migration paths: from China to Japan ($p = 0.035$), from China to the USA ($p = 0.0030$), and from the USA to Europe ($p = 0.0017$). The relative entropy data, therefore, confirms China as the H3N2 reservoir and implies that novel H3N2 viruses are emerging in China, not during the migration process.

\subsection{Positions under Selection}
\label{sec:Positions_under_Selection}

\begin{figure}
\centering
\vspace{-1in}
\subfigure[]{
\includegraphics[width=2.75in]{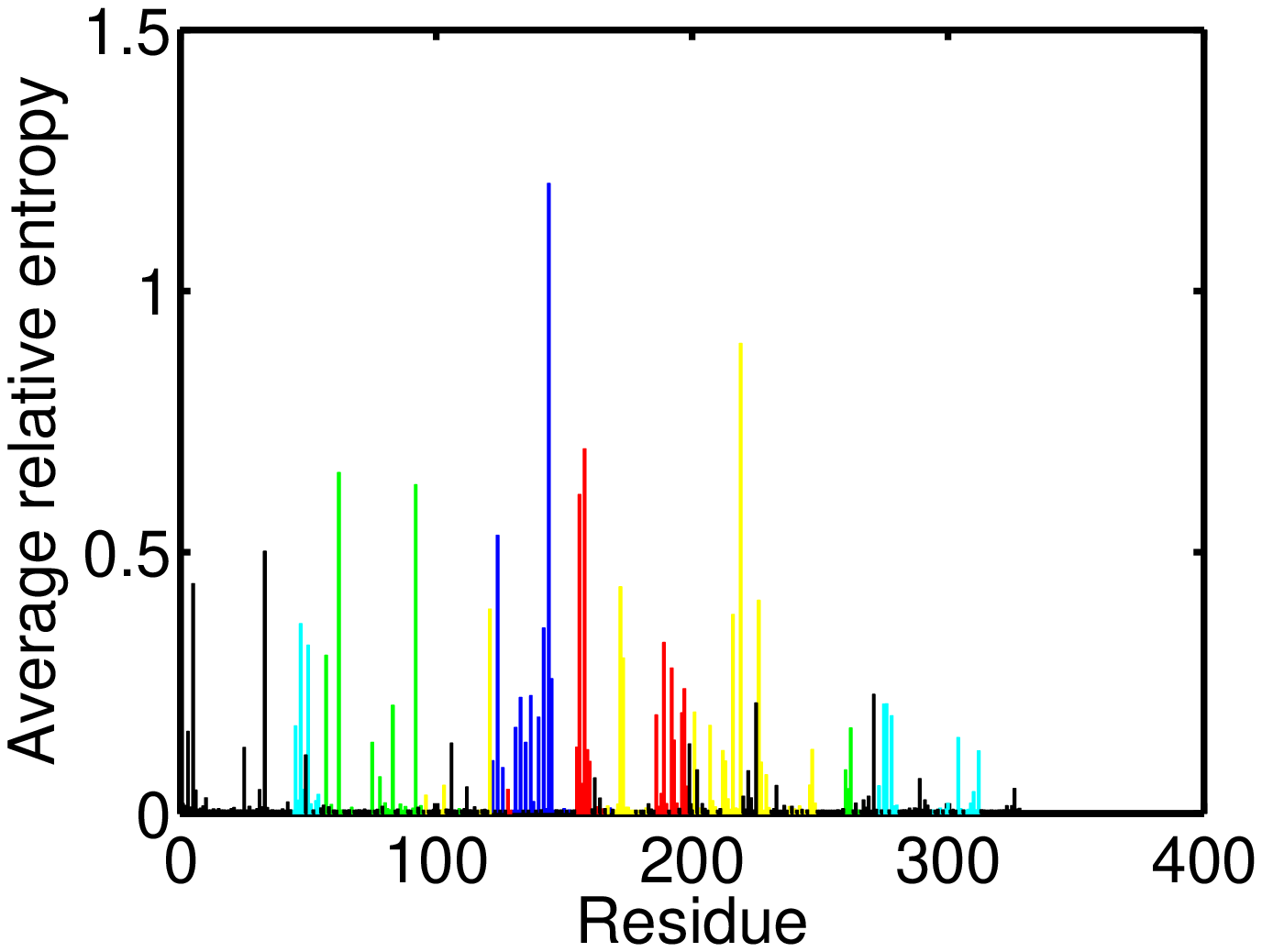}
\label{fig:avg_select}
}
\subfigure[]{
\includegraphics[width=2.75in]{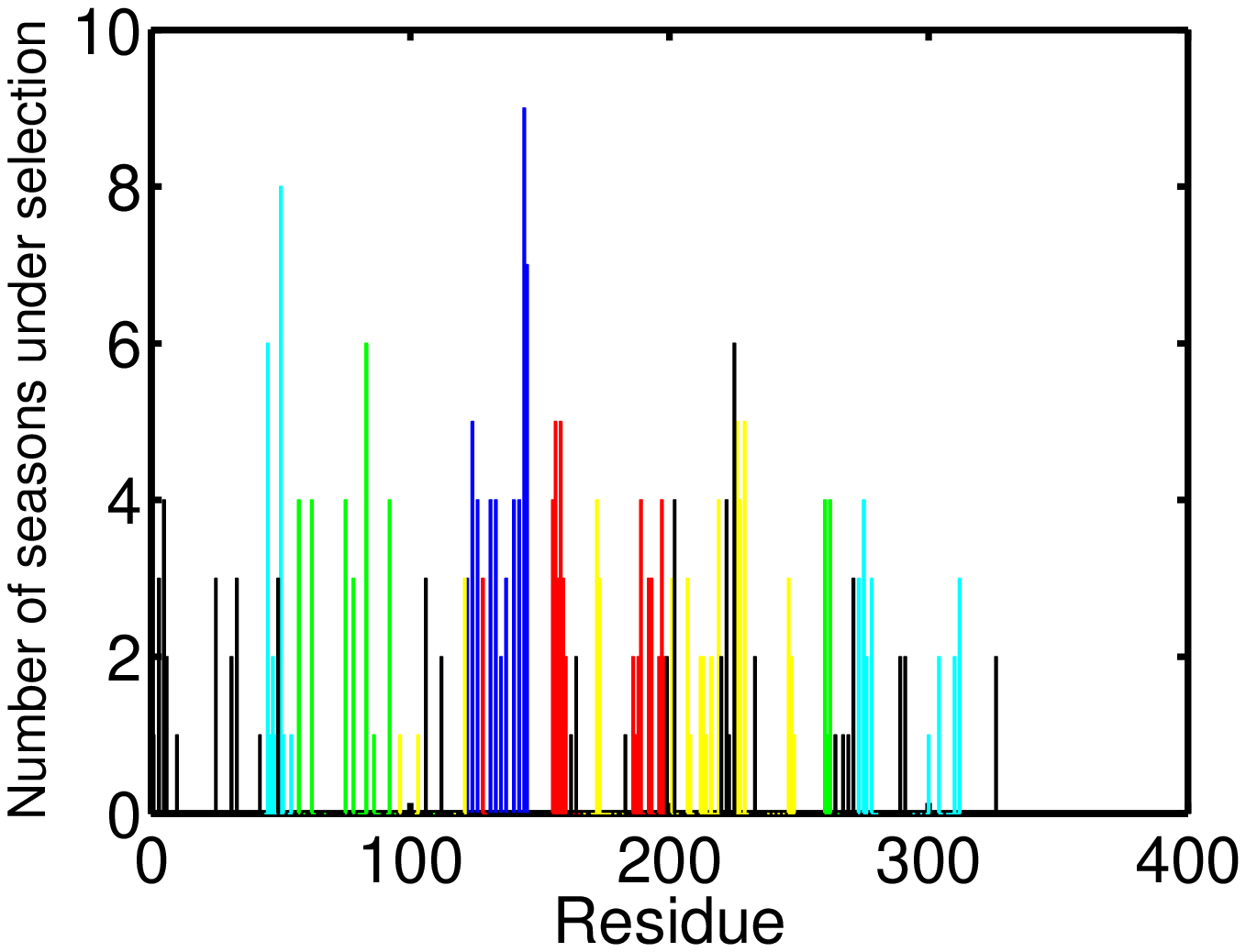}
\label{fig:n_select}
}
\subfigure[]{
\includegraphics[width=2.75in]{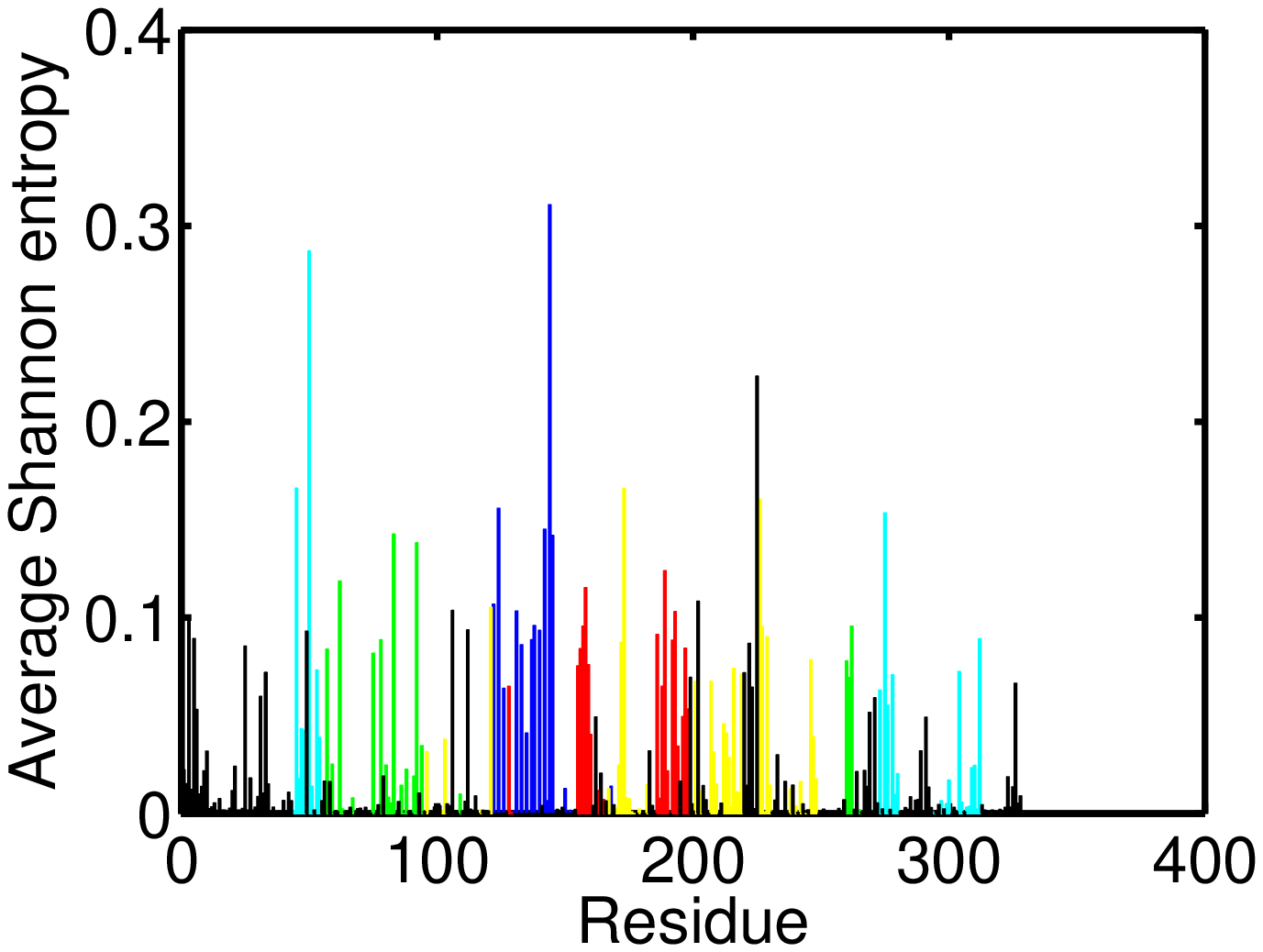}
\label{fig:avg_div}
}
\subfigure[]{
\includegraphics[width=2.75in]{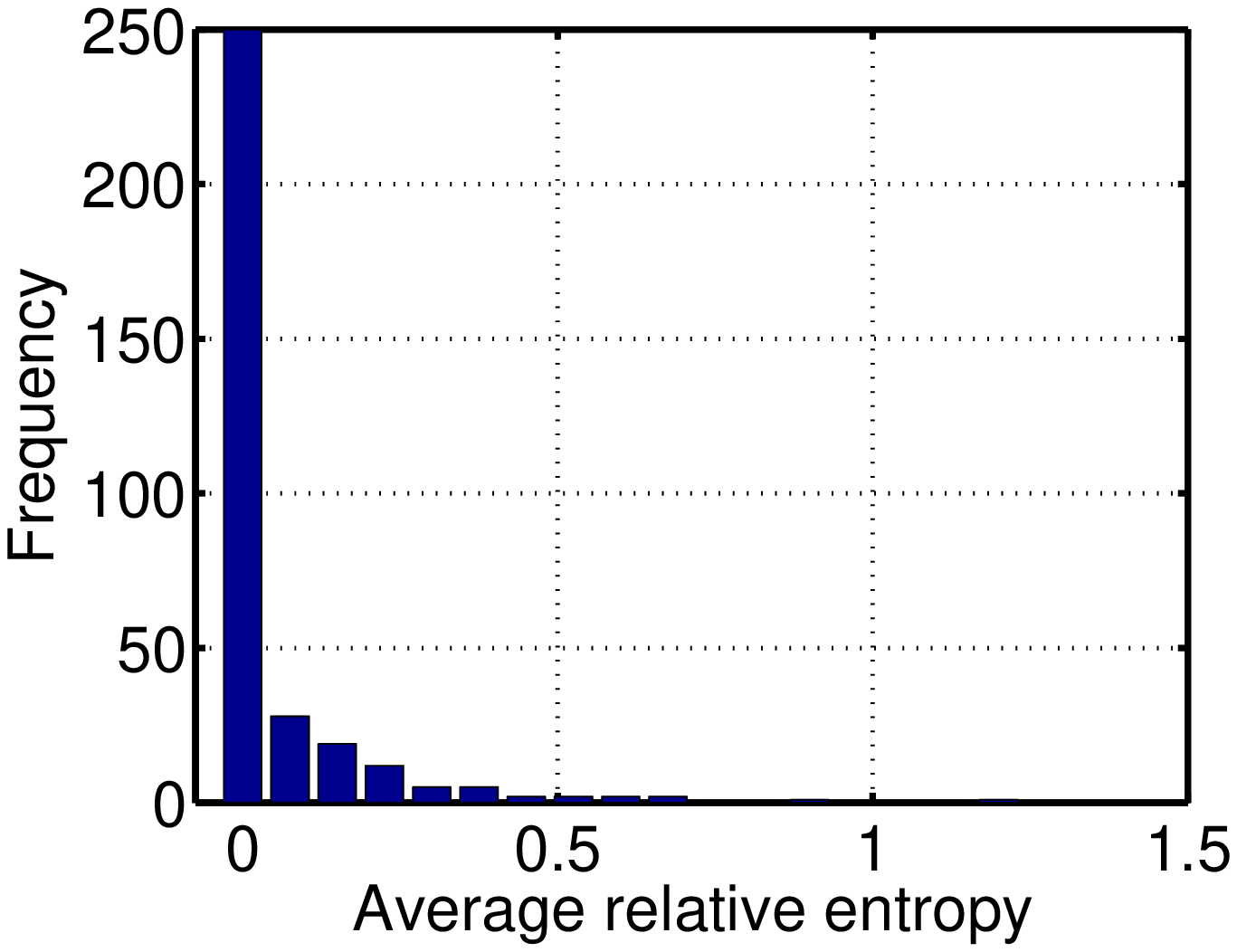}
\label{fig:dist_avg_select}
}
\subfigure[]{
\includegraphics[width=2.75in]{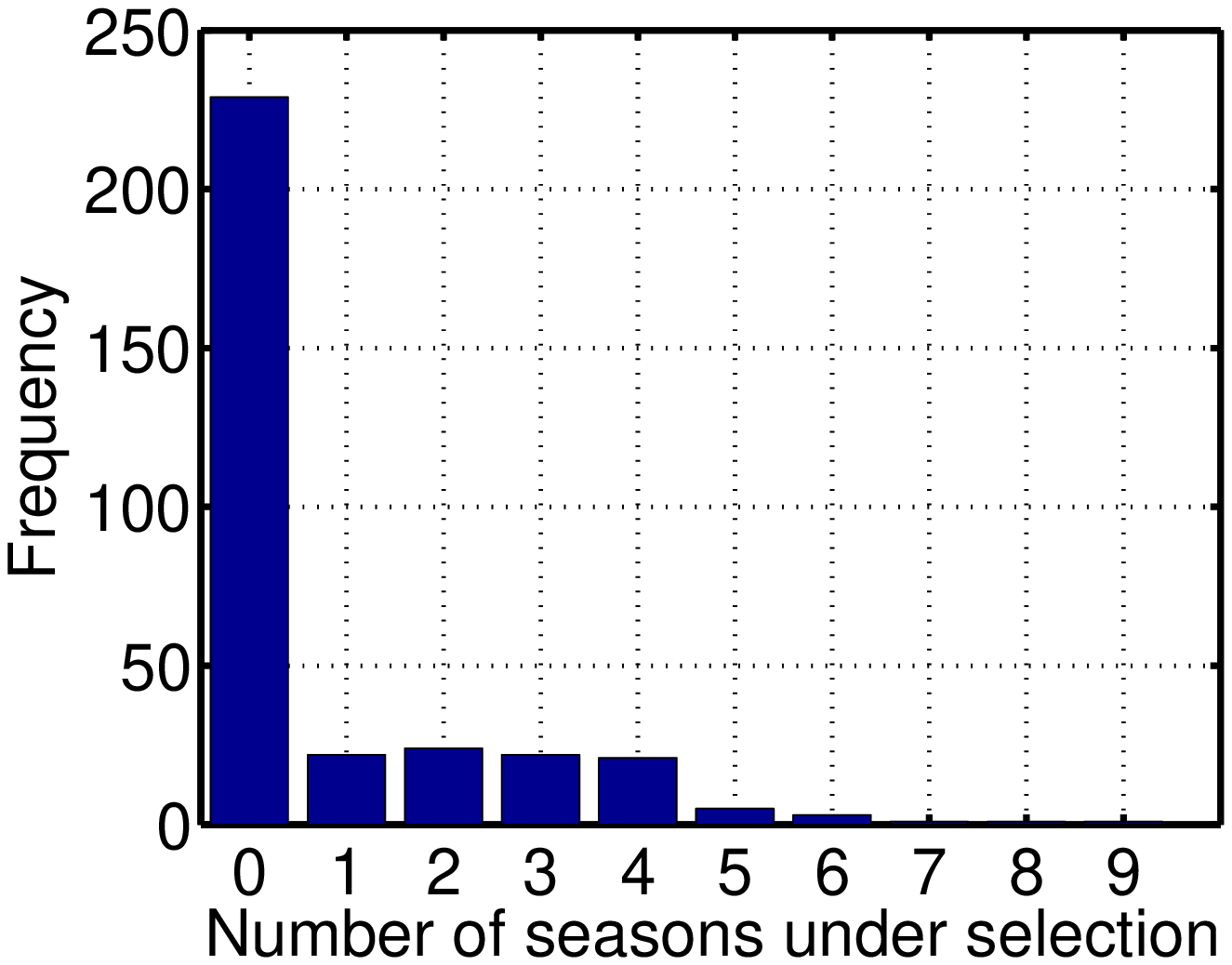}
\label{fig:dist_n_select}
}
\subfigure[]{
\includegraphics[width=2.75in]{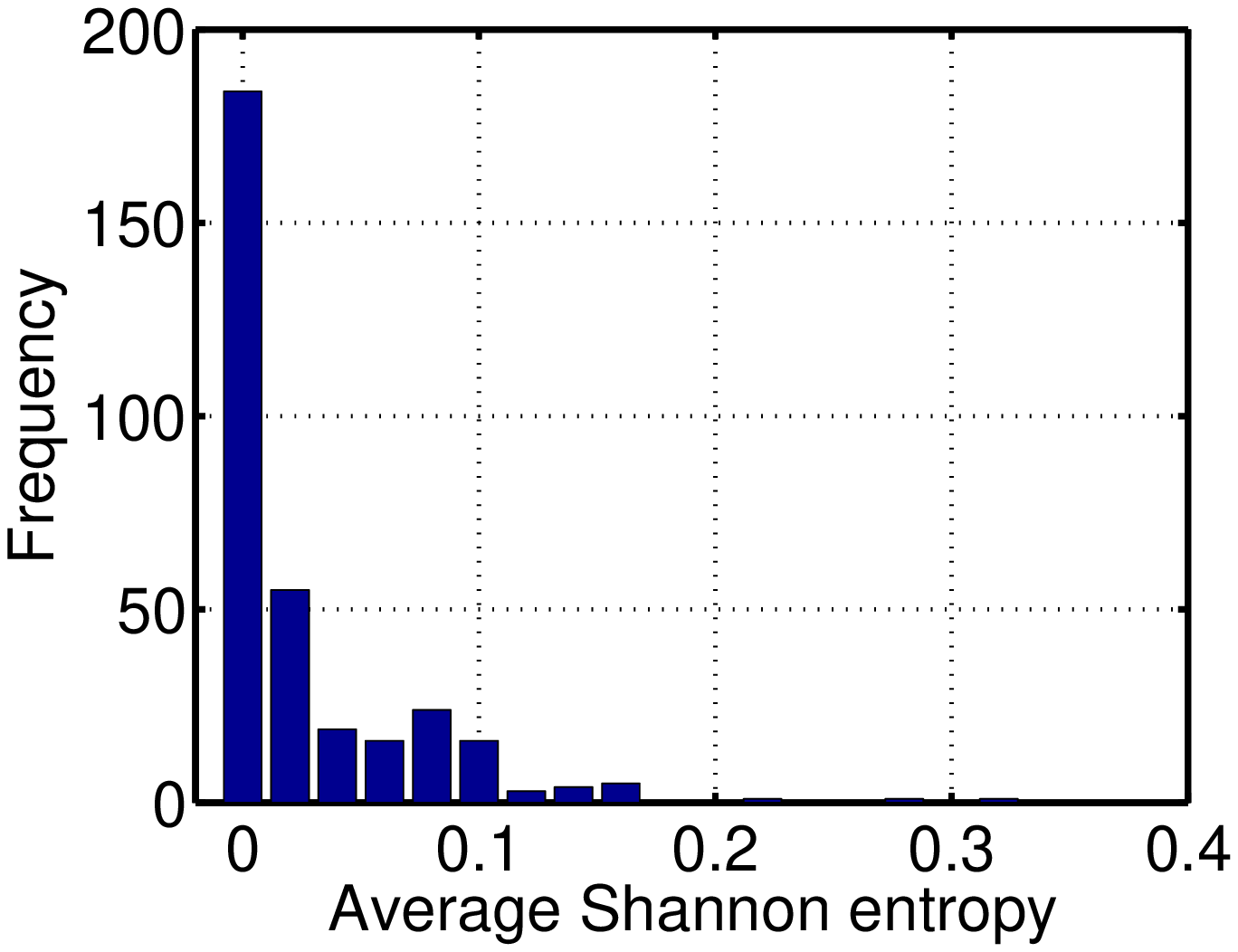}
\label{fig:dist_avg_div}
}
\caption{\subref{fig:avg_select} Average selection in each position quantified by relative entropy during the past 17 seasons from 1993--1994 to 2009--2010, calculated by $\bar{S}_j = \sum_{i=1}^{17} S_{i,j} / 17$. The colors represent positions in epitopes A to E and positions outside the epitopes, as in Figure \ref{fig:H3_epitope}. \subref{fig:n_select} Number of seasons for each position when the relative entropy was greater than the threshold $S_i^\mathrm{thres}$, i.e. the position was under selection. \subref{fig:avg_div} Average diversity in each position quantified by Shannon entropy in the seasons from 1993--1994 to 2009--2010, calculated by $\bar{D}_j = \sum_{i=1}^{17} D_{i,j} / 17$. \subref{fig:dist_avg_select} Distribution of the average selection in each position displayed in \subref{fig:avg_select}. \subref{fig:dist_n_select} Distribution of the numbers of seasons under selection displayed in \subref{fig:n_select}. \subref{fig:dist_avg_div} Distribution of the average diversity in each position shown in \subref{fig:avg_div}.}
\label{fig:spatial}
\end{figure}

The values of diversity as Shannon entropy $D_{i,j}$ and selection as relative entropy $S_{i,j}$ are available for the sequences collected from the 1993--1994 season to the 2009--2010 season. First we apply the mean field approximation to remove the variation of selection and diversity over the time, and only consider the variation of Shannon entropy and relative entropy in different positions and regions over the past 17 seasons. A profile of the pattern of Shannon entropy and relative entropy in position $j$ comprises the average selection, the number of seasons under selection, and the average diversity. The average selection $\bar{S}_j$
is expressed by the mean of relative entropy in each position over the 17 seasons
\begin{equation}
\bar{S}_j = \frac{1}{17}\sum_{i=1}^{17} S_{i,j}
\end{equation}
and is displayed in Figure \ref{fig:spatial} \subref{fig:avg_select}. The number of seasons under selection in each position $j$ is calculated by
\begin{equation}\label{eq:n_select}
N_j = \sum_{i=1}^{17} H\left(S_{i,j} - S_i^\mathrm{thres}\right)
\end{equation}
where $H$ is the Heaviside step function. The numbers are shown in Figure \ref{fig:spatial} \subref{fig:n_select}. The average diversity $\bar{D}_j$ in each position is calculated by averaging the Shannon entropy over the 17 seasons from 1993--1994 to 2009--2010
\begin{equation}
\bar{D}_j = \frac{1}{17}\sum_{i=1}^{17} D_{i,j}
\end{equation}
and is displayed in Figure \ref{fig:spatial} \subref{fig:avg_div}.

Figure \ref{fig:spatial} \subref{fig:dist_avg_select} presents the distribution of the selection.  Around 76\% of the amino acid positions 1--329 of hemagglutinin have an average selection close to zero and fall into the leftmost bin.  The numbers of seasons when selection $S_{i,j}$ in these positions were greater than the threshold level $S_i^\mathrm{thres}$ are shown in Figure \ref{fig:spatial} \subref{fig:dist_n_select}.  The average diversities $\bar{D}_{i,j}$ in all the positions are shown in Figure \ref{fig:spatial} \subref{fig:dist_avg_div}.  If position $j$ is under selection with $S_{i,j} > S_i^\mathrm{thres}$ in greater than two of the 17 seasons between 1993--1994 and 2009--2010, or $N_j > 2$, this position $j$ is classified as a position under selection in the evolutionary history of H3N2 virus.  The 54 positions with $S_{i,j} > S_i^\mathrm{thres}$ in greater than two seasons are listed in Table \ref{tab:positions}.

Patterns of selection and diversity similar to those observed in historical sequences in Figure \ref{fig:spatial} are generated by a Monte Carlo simulation model, as displayed in Figure S1.  The basis of the Monte Carlo simulation is that antibody binds to one of two epitope regions on the surface of the HA1 domain \citep{Gupta2006}, and the dominant epitope bound by antibody is under immune pressure and undergoes a higher substitution rate \citep{Ferguson2003}.  The detailed description and discussion of the Monte Carlo model is in the Appendix.

\begin{table}
\caption{Amino acid positions $j$ under selection. To be included, the positions must be under selection, $S_{i,j} > S_i^\mathrm{thres}$, in greater than two seasons.}\centering
\begin{tabular}{l l}
\\\hline
Region          & Amino acid positions \\\hline
Epitope A       & 122   124   126   131   133   137   140   142   144   145 \\
Epitope B       & 128   155   156   157   158   159   189   192   193   197 \\
Epitope C       &  45    50   273   275   278   312 \\
Epitope D       & 121   172   173   201   207   219   226   227   229   246 \\
Epitope E       &  57    62    75    78    83    92   260   262 \\
Out of epitopes &   3     5    25    33    49   106   202   222   225   271 \\\hline
\end{tabular}
\label{tab:positions}
\end{table}

\subsection{Comparison of Different Regions}
\label{sec:Comparison_of_Different_Regions}

A human antibody binds to five epitopes in the H3 hemagglutinin \citep{Wiley1981}. The five epitopes are located in different parts of the HA1 domain of the cylinder-like structure of the H3 hemagglutinin. Epitopes A and B are on the top of the HA1 structure and are exposed in the HA trimer. Epitope D is on the top of HA1 structure and is partly buried inside the HA trimer. Epitopes C and E are at the central area of the exposed surface of the HA1 domain as shown in Figure \ref{fig:H3_epitope}. Using the entropy method, we will show that epitopes A and B are under the highest average selection over all the seasons. These results can be interpreted as the antibody binds mostly to the top exposed part of the structure of hemagglutinin trimer defined by epitopes A and B, and so the selection in these two epitopes is with higher intensity.

We divide the HA1 domain of the H3N2 hemagglutinin into six regions, namely epitopes A to E, and positions not in any of the epitopes. These regions show significantly distinct patterns of evolution. In each seasons from 1993--1994 to 2009--2010, we averaged selection and diversity in each epitope and the positions not in any of the epitopes. The fraction of positions $j$ under selection defined by $S_{i,j} > S_i^\mathrm{thres}$ was also calculated. The averages for 17 seasons are listed in Table \ref{tab:epitopes}. It is evident that the values in Table \ref{tab:epitopes} vary across the epitopes. The selection and diversity in epitopes A and B are greater than those in epitopes C, D, and E for each of selection, fraction of positions under selection, and diversity. The fraction of positions under selection is significantly greater than those in epitopes C, D, and E ($p < 0.038$, using Wilcoxon signed rank test). The values in epitopes C, D, and E are significantly greater than those not in any of the epitopes ($p < 0.0019$ for selection, $p < 0.0011$ for fraction of positions under selection, and $p < 6.0 \times 10^{-4}$ for diversity, using Wilcoxon signed rank test). Consequently, epitopes A and B display the highest level of selection and diversity.

\begin{table}
\caption{Annual selection, fraction of positions under selection, and diversity in epitopes A to E, positions not in any of the epitopes, and the whole HA1 sequence.} \centering
\begin{tabular}{l l l l}
\\\hline
Region             & Selection & Fraction of positions & Diversity \\
                   &           & under selection       &           \\\hline
Epitope A          & 0.187 & 0.152 & 0.077 \\
Epitope B          & 0.157 & 0.134 & 0.062 \\
Epitope C          & 0.077 & 0.087 & 0.048 \\
Epitope D          & 0.100 & 0.072 & 0.037 \\
Epitope E          & 0.111 & 0.094 & 0.049 \\
Out of epitopes    & 0.021 & 0.019 & 0.013 \\
The whole sequence & 0.060 & 0.051 & 0.028 \\\hline
\end{tabular}
\label{tab:epitopes}
\end{table}

\section{Discussion}
\label{sec:Discussion}

The Shannon entropy and the relative entropy are here introduced to quantify the diversity and the selection pressure of the evolving H3N2 virus. The foundation of the entropy calculation is the assumption that the virus sequences used in the entropy calculation are from a random unbiased sampling of the virus circulating in the human population. However, sampling density of the H3N2 virus varies in different geographic regions, hence creating a sampling bias.

We have addressed this issue at the continent level by analyzing data from different regions separately. We now additionally study the effect of sampling bias within one country. For example, among the H3 hemagglutinin sequences labeled with month of collection in the NCBI Influenza Virus Resource Database, the New York state sequences account for about one third of the USA sequences. In the contrast, New York state has only about 6.5\% of the USA population. We chose eight seasons from 2001--2002 to 2008--2009 with abundant USA sequences collected in and out of New York state during this period of time available in the database. Using the procedure in the Materials and Methods section, we calculated the histogram of 20 amino acids in each amino acid position in each season for the New York state sequences, and that for the non-New York state sequences. Each of the $329 \times 8 = 2632$ pairs of histograms in and out of New York state were compared using the $\chi^2$ test for homogeneity. The $p$ values of 2581 pairs are greater than 0.05. That is, 98.1\% of the pairs are not significantly different. The high sampling density in New York state does not affect the histograms of 20 amino acids in each position, implying that the sampling bias of the H3N2 virus is uncorrelated to the amino acid usage patterns.

By applying Shannon entropy and relative entropy to the aligned hemagglutinin sequences labeled with month of collection, we obtain the evolution and migration pattern of the H3N2 virus in the Results section. First, Shannon entropy and relative entropy quantify diversity of and selection pressure over the virus, relatively. Relative entropy from the current season $i$ to the next season $i+1$ linearly increased with the Shannon entropy in the current season $i$. See Figure \ref{fig:entropy_allposition} and \ref{fig:entropy_avg_year}. Second, relative entropy quantifies the similarity of two groups of virus and implies the migration path of the H3N2 virus. See Table \ref{tab:migration}. In the following text we compare our methods and results to the literature.

The relative entropy reveals the H3N2 migration pattern. Previous studies applied phylogenetic methods in an attempt to locate the epicenter of the H3N2 epidemic in each season. \citet{Rambaut2008} studied the dynamic of influenza sequence diversity in the temperate regions in both hemispheres to imply that the H3N2 virus originates in the tropics and migrates to the temperate regions in both hemispheres. \citet{Russell2008} obtained the antigenic and genetic evolution rate in each region and distances of the H3N2 strains to the trunk of the phylogenic tree. This information indicated East and Southeast Asia as the epicenter, from which the H3N2 virus spreads to North America, Europe, and Oceania in each season \citep{Russell2008}. Recently, \citet{Bedford2010} suggested the center of the H3N2 migration network being China, Southeast Asia, and the USA by estimating the migration rate between different regions in the world. Here the relative entropy, the gain of sequence information, is used as a novel measure of the sequence similarity. The H3N2 migration path is the directed graph in which each path has the minimum relative entropy, or the maximum sequence similarity. These studies reach a consensus that South China is located in the epicenter of influenza epidemics. Here, we additionally identify a novel migration path from the USA to Europe and show that virus evolutionary rate is higher in the epicenter than in the migration paths.

Previous studies have identified positions that have led to the immune escape of influenza, by resolving historical mutations. A 1997 study examining 254 H3 nucleotide sequences from 1984 to 1996 identified 14 positions that are under selection, using the dN/dS ratio method \citep{Fitch1997}. In a 1999 study involving 357 nucleotide sequences from 1983 to 1997, 18 positions were identified to be under selection, using the dN/dS ratio method \citep{Bush1999}. A 2003 paper used the alignment of 525 nucleotide sequences from 1968 to 2000 to calculate the codon diversity and the amino acid diversity \citep{Plotkin2003}. This paper reported 25 positions with the largest codon diversity and amino acid diversity. A 2007 paper located 63 positions of positive selection by alignment of 2,248 sequences from 1968 to 2005 and considering substitutions at the amino acid level \citep{Shih2007}. Our study identified 54 positions under selection at the amino acid level, using 4,292 aligned sequences from the 1992--1993 season to the 2009--2010 season. Considering this historical body of work, it is apparent that the number of amino acid positions identified to be under selection has increased with the number and the time span of sequences used in the study and as the discriminating power of the data has increased. In addition, different criteria to identify the positions under selection have been introduced in the previous studies \citep{Fitch1997,Bush1999,Plotkin2003,Shih2007} and in the present study. \citet{Shih2007} identified positions to be under selection when an amino acid substitution occurred during successive years. We here classify a position $j$ as under selection if its relative entropy $S_{i,j}$ is greater than the threshold $S_i^\mathrm{thres}$ in greater than two seasons. These two methods identify many identical positions as well as some distinct positions.

The heterogeneity of the methods also contributes to identification of different sets of positions under selection. We note that these methods fall into two categories. The first category operates at the codon level. The $\mathrm{d}N/\mathrm{d}S$ ratio method \citep{Fitch1997,Bush1999} calculates the non-synonymous and synonymous mutation rate of the codon. The \citet{Plotkin2003} method comparing the codon diversity and the amino acid diversity is a variation of the $\mathrm{d}N/\mathrm{d}S$ ratio method. The second category operates at the amino acid level. \citet{Shih2007} identified the positions with amino acid switch occurring in history to be under selection. Our entropy method recognizes the positions with relative entropy higher than the threshold, $S_i^\mathrm{thres}$, in greater than two seasons. A large $\mathrm{d}N/\mathrm{d}S$ of a codon does not necessarily mean an amino acid switch in the same position because the amino acid substitution could be unfixed. Methods at the amino acid level, such as the amino acid switch \citep{Shih2007} and our entropy method, can identify positions with low $\mathrm{d}N/\mathrm{d}S$ to be under selection because these methods do not consider nucleotide substitutions. Positive selection does not necessarily lead to a fixed amino acid switch, and in this case the entropy method can still detect positive selection. Unlike the amino acid switch method \citep{Shih2007}, the entropy method applied in this study is able to detect unfixed amino acid substitutions arising from selection. Our entropy method releases the requirement of fixed amino acid substitution in \citep{Shih2007} but adds one requirement: the positions under selection need to present large relative entropy in greater than two seasons. Consequently, these methods identify slightly different sets of positions to be under selection.

\section{Conclusion}
\label{sec:Conclusion}

We use Shannon entropy and relative entropy as two state variables of H3N2 evolution. The entropy method is able to predict H3N2 evolution and migration in the next season. First, we show that the rate of evolution increases with the virus diversity in the current season. The Shannon entropy data in one season strongly correlate with the relative entropy data from that season to the next season. If higher Shannon entropy of the virus is observed in one season, higher virus evolutionary rate is expected from this season to the next season. Second, the relative entropy values between virus sequences from China, Japan, the USA, and Europe indicate that the H3N2 virus migration from China to Japan and the USA, and identify a novel migration path from the USA and Europe. The relative entropy values in and out of China, the epicenter, show that evolutionary rate is higher in China than in the migration paths. Moreover, the entropy method was demonstrated on two applications. First, selection pressure of the H3 hemagglutinin is mainly in 54 amino acid positions. Second, the top exposed part in the three-dimensional structure of HA trimer covered by epitopes A and B is under the highest level of selection. These results substantiate current thinking on H3N2 evolution, and show that the selection pressure is focused in a subset of amino acid positions in the epitopes, with epitopes A and B on the top of hemagglutinin being dominant and making the largest contribution to the H3N2 evolution. These predictions and applications show that the entropy method is not only predictive but also descriptive.

\paragraph{Acknowledgements} Keyao Pan's research was supported by a training fellowship from the Keck Center Nanobiology Training Program of the Gulf Coast Consortia (NIH Grant No.\ R90 DK071504).  This project was also partially supported by DARPA grant HR 0011-09-1-0055.

\newpage

\bibliographystyle{rspublicnat}
\bibliography{references}

\begin{thebibliography}{3}
\providecommand{\natexlab}[1]{#1}
\expandafter\ifx\csname urlstyle\endcsname\relax
  \providecommand{\doi}[1]{doi:\discretionary{}{}{}#1}\else
  \providecommand{\doi}{doi:\discretionary{}{}{}\begingroup
  \urlstyle{rm}\Url}\fi

\bibitem[{Ferguson \emph{et~al.}(2003)Ferguson, Galvani \& Bush}]{Ferguson2003}
Ferguson, N.~M., Galvani, A.~P. \& Bush, R.~M. 2003 Ecological and
  immunological determinants of influenza evolution.
\newblock \emph{Nature}, \textbf{422}, 428--433.
\newblock (\doi{10.1038/nature01509})

\bibitem[{Gupta \emph{et~al.}(2006)Gupta, Earl \& Deem}]{Gupta2006}
Gupta, V., Earl, D.~J. \& Deem, M.~W. 2006 Quantifying influenza vaccine
  efficacy and antigenic distance.
\newblock \emph{Vaccine}, \textbf{24}, 3881--3888.
\newblock (\doi{10.1016/j.vaccine.2006.01.010})

\bibitem[{Shih \emph{et~al.}(2007)Shih, Hsiao, Ho \& Li}]{Shih2007}
Shih, A.~C., Hsiao, T.~C., Ho, M.~S. \& Li, W.~H. 2007 Simultaneous amino acid
  substitutions at antigenic sites drive influenza {A} hemagglutinin evolution.
\newblock \emph{Proc. Natl. Acad. Sci. USA}, \textbf{104}, 6283--6288.
\newblock (\doi{10.1073/pnas.0701396104})

\end{thebibliography}


\begin{thebibliography}{36}
\providecommand{\natexlab}[1]{#1}
\expandafter\ifx\csname urlstyle\endcsname\relax
  \providecommand{\doi}[1]{doi:\discretionary{}{}{}#1}\else
  \providecommand{\doi}{doi:\discretionary{}{}{}\begingroup
  \urlstyle{rm}\Url}\fi

\bibitem[{WHO()}]{WHO2009}
 {W}orld {H}ealth {O}rganization {M}edia {C}entre influenza fact sheet 211.
\newblock \url{http://www.who.int/mediacentre/factsheets/fs211/en/index.html}.

\bibitem[{Bedford \emph{et~al.}(2010)Bedford, Cobey, Beerli \&
  Pascual}]{Bedford2010}
Bedford, T., Cobey, S., Beerli, P. \& Pascual, M. 2010 Global migration
  dynamics underlie evolution and persistence of human influenza {A}
  ({H}3{N}2).
\newblock \emph{PLoS Pathog.}, \textbf{6}, e1000\,918.
\newblock (\doi{10.1371/journal.ppat.1000918})

\bibitem[{Bush \emph{et~al.}(1999)Bush, Fitch, Bender \& Cox}]{Bush1999}
Bush, R.~M., Fitch, W.~M., Bender, C.~A. \& Cox, N.~J. 1999 Positive selection
  on the {H}3 hemagglutinin gene of human influenza virus {A}.
\newblock \emph{Mol. Biol. Evol.}, \textbf{16}, 1457--1465.

\bibitem[{Deem \& Lee(2003)}]{Deem2003}
Deem, M.~W. \& Lee, H.~Y. 2003 Sequence space localization in the immune system
  response to vaccination and disease.
\newblock \emph{Phys. Rev. Lett.}, \textbf{91}, 068\,101.
\newblock (\doi{10.1103/PhysRevLett.91.068101})

\bibitem[{Deem \& Pan(2009)}]{Deem2009}
Deem, M.~W. \& Pan, K. 2009 The epitope regions of {H}1--subtype influenza {A},
  with application to vaccine efficacy.
\newblock \emph{Protein Eng., Des. Sel.}, \textbf{22}, 543--546.
\newblock (\doi{10.1093/protein/gzp027})

\bibitem[{Ferguson \emph{et~al.}(2003)Ferguson, Galvani \& Bush}]{Ferguson2003}
Ferguson, N.~M., Galvani, A.~P. \& Bush, R.~M. 2003 Ecological and
  immunological determinants of influenza evolution.
\newblock \emph{Nature}, \textbf{422}, 428--433.
\newblock (\doi{10.1038/nature01509})

\bibitem[{Fitch \emph{et~al.}(1997)Fitch, Bush, Bender \& Cox}]{Fitch1997}
Fitch, W.~M., Bush, R.~M., Bender, C.~A. \& Cox, N.~J. 1997 Long term trends in
  the evolution of {H}(3) {HA}1 human influenza type {A}.
\newblock \emph{Proc. Natl. Acad. Sci. USA}, \textbf{94}, 7712--7718.
\newblock (\doi{10.1073/pnas.94.15.7712})

\bibitem[{Gerstein \& Altman(1995)}]{Gerstein1995}
Gerstein, M. \& Altman, R.~B. 1995 Average core structures and variability
  measures for protein families: Application to the immunoglobulins.
\newblock \emph{J. Mol. Biol.}, \textbf{251}, 161--175.
\newblock (\doi{10.1006/jmbi.1995.0423})

\bibitem[{Gupta \emph{et~al.}(2006)Gupta, Earl \& Deem}]{Gupta2006}
Gupta, V., Earl, D.~J. \& Deem, M.~W. 2006 Quantifying influenza vaccine
  efficacy and antigenic distance.
\newblock \emph{Vaccine}, \textbf{24}, 3881--3888.
\newblock (\doi{10.1016/j.vaccine.2006.01.010})

\bibitem[{Halabi \emph{et~al.}(2009)Halabi, Rivoire, Leibler \&
  Ranganathan}]{Halabi2009}
Halabi, N., Rivoire, O., Leibler, S. \& Ranganathan, R. 2009 Protein sectors:
  Evolutionary units of three-dimensional structure.
\newblock \emph{Cell}, \textbf{138}, 774--786.
\newblock See the supplemental data for the derivation of Equation 7.
\newblock (\doi{10.1016/j.cell.2009.07.038})

\bibitem[{Ina \& Gojobori(1994)}]{Ina1994}
Ina, Y. \& Gojobori, T. 1994 Statistical analysis of nucleotide sequences of
  the hemagglutinin gene of human influenza {A} viruses.
\newblock \emph{Proc. Natl. Acad. Sci. USA}, \textbf{91}, 8388--8392.
\newblock (\doi{10.1073/pnas.91.18.8388})

\bibitem[{Kullback \& Leibler(1951)}]{Kullback1951}
Kullback, S. \& Leibler, R.~A. 1951 On information and sufficiency.
\newblock \emph{Ann. Math. Statist.}, \textbf{22}, 79--86.
\newblock (\doi{10.1214/aoms/1177729694})

\bibitem[{Li \emph{et~al.}(2007)Li, Carroll, Gardner, Walsh, Vitalis \&
  Damon}]{Li2007}
Li, Y., Carroll, D.~S., Gardner, S.~N., Walsh, M.~C., Vitalis, E.~A. \& Damon,
  I.~K. 2007 On the origin of smallpox: {C}orrelating variola phylogenics with
  historical smallpox records.
\newblock \emph{Proc. Natl. Acad. Sci. USA}, \textbf{104}, 15\,787--15\,792.
\newblock (\doi{10.1073/pnas.0609268104})

\bibitem[{Liao \emph{et~al.}(2008)Liao, Lee, Ko \& Hsiung}]{Liao2008}
Liao, Y.~C., Lee, M.~S., Ko, C.~Y. \& Hsiung, C.~A. 2008 Bioinformatics models
  for predicting antigenic variants of influenza {A}/{H}3{N}2 virus.
\newblock \emph{Bioinformatics}, \textbf{24}, 505--512.
\newblock (\doi{10.1093/bioinformatics/btm638})

\bibitem[{Lockless \& Ranganathan(1999)}]{Lockless1999}
Lockless, S.~W. \& Ranganathan, R. 1999 Evolutionarily conserved pathways of
  energetic connectivity in protein families.
\newblock \emph{Science}, \textbf{286}, 295--299.
\newblock (\doi{10.1126/science.286.5438.295})

\bibitem[{Mirny \& Shakhnovich(1999)}]{Mirny1999}
Mirny, L.~A. \& Shakhnovich, E.~I. 1999 Universally conserved positions in
  protein folds: Reading evolutionary signals about stability, folding kinetics
  and function.
\newblock \emph{J. Mol. Biol.}, \textbf{291}, 177--196.
\newblock (\doi{10.1006/jmbi.1999.2911})

\bibitem[{Nobusawa \& Sato(2006)}]{Nobusawa2006}
Nobusawa, E. \& Sato, K. 2006 Comparison of the mutation rates of human
  influenza {A} and {B} viruses.
\newblock \emph{J. Virol.}, \textbf{80}, 3675--3678.
\newblock The average mutation rate of influenza A virus is equivalent to $1.6
  \times 10^{-5}/\mathrm{residue}/\mathrm{day}$, or $5.8 \times
  10^{-3}/\mathrm{residue}/\mathrm{year}$.
\newblock (\doi{10.1128/JVI.80.7.3675-3678.2006})

\bibitem[{Parvin \emph{et~al.}(1986)Parvin, Moscona, Pan, Leider \&
  Palese}]{Parvin1986}
Parvin, J.~D., Moscona, A., Pan, W.~T., Leider, J.~M. \& Palese, P. 1986
  Measurement of the mutation rates of animal viruses: Influenza {A} virus and
  poliovirus type 1.
\newblock \emph{J. Virol.}, \textbf{59}, 377--383.

\bibitem[{Plaxco \emph{et~al.}(2000)Plaxco, Larson, Ruczinski, Riddle, Thayer,
  Buchwitz, Davidson \& Baker}]{Plaxco2000}
Plaxco, K.~W., Larson, S., Ruczinski, I., Riddle, D.~S., Thayer, E.~C.,
  Buchwitz, B., Davidson, A.~R. \& Baker, D. 2000 Evolutionary conservation in
  protein folding kinetics.
\newblock \emph{J. Mol. Biol.}, \textbf{298}, 303--312.
\newblock (\doi{10.1006/jmbi.1999.3663})

\bibitem[{Plotkin \& Dushoff(2003)}]{Plotkin2003}
Plotkin, J.~B. \& Dushoff, J. 2003 Codon bias and frequency-dependent selection
  on the hemagglutinin epitopes of influenza {A} virus.
\newblock \emph{Proc. Natl. Acad. Sci. USA}, \textbf{100}, 7152--7157.
\newblock (\doi{10.1073/pnas.1132114100})

\bibitem[{Plotkin \emph{et~al.}(2002)Plotkin, Dushoff \& Levin}]{Plotkin2002}
Plotkin, J.~B., Dushoff, J. \& Levin, S.~A. 2002 Hemagglutinin sequence
  clusters and the antigenic evolution of influenza {A} virus.
\newblock \emph{Proc. Natl. Acad. Sci. USA}, \textbf{99}, 6263--6268.
\newblock (\doi{10.1073/pnas.082110799})

\bibitem[{Rambaut \emph{et~al.}(2008)Rambaut, Pybus, Nelson, Viboud,
  Taubenberger \& Holmes}]{Rambaut2008}
Rambaut, A., Pybus, O.~G., Nelson, M.~I., Viboud, C., Taubenberger, J.~K. \&
  Holmes, E.~C. 2008 The genomic and epidemiological dynamics of human
  influenza {A} virus.
\newblock \emph{Nature}, \textbf{453}, 615--U2.
\newblock (\doi{10.1038/nature06945})

\bibitem[{Russell \emph{et~al.}(2008)Russell, Jones, Barr, Cox, Garten,
  Gregory, Gust, Hampson, Hay \emph{et~al.}}]{Russell2008}
Russell, C.~A., Jones, T.~C., Barr, I.~G., Cox, N.~J., Garten, R.~J., Gregory,
  V., Gust, I.~D., Hampson, A.~W., Hay, A.~J. \emph{et~al.} 2008 The global
  circulation of seasonal influenza {A} ({H}3{N}2) viruses.
\newblock \emph{Science}, \textbf{320}, 340--346.
\newblock (\doi{10.1126/science.1154137})

\bibitem[{Sander \& Schneider(1991)}]{Sander1991}
Sander, C. \& Schneider, R. 1991 Database of homology-derived protein
  structures and the structural meaning of sequence alignment.
\newblock \emph{Proteins}, \textbf{9}, 56--68.
\newblock (\doi{10.1002/prot.340090107})

\bibitem[{Schneider \& Stephens(1990)}]{Schneider1990}
Schneider, T.~D. \& Stephens, R.~M. 1990 Sequence logos: a new way to display
  consensus sequences.
\newblock \emph{Nucleic Acids Res.}, \textbf{18}, 6097--6100.
\newblock (\doi{10.1093/nar/18.20.6097})

\bibitem[{Schneider \emph{et~al.}(1986)Schneider, Stormo, Gold \&
  Ehrenfeucht}]{Schneider1986}
Schneider, T.~D., Stormo, G.~D., Gold, L. \& Ehrenfeucht, A. 1986 Information
  content of binding sites on nucleotide sequences.
\newblock \emph{J. Mol. Biol.}, \textbf{188}, 415--431.
\newblock (\doi{10.1016/0022-2836(86)90165-8})

\bibitem[{Shenkin \emph{et~al.}(1991)Shenkin, Erman \&
  Mastrandrea}]{Shenkin1991}
Shenkin, P.~S., Erman, B. \& Mastrandrea, L.~D. 1991 Information-theoretical
  entropy as a measure of sequence variability.
\newblock \emph{Proteins}, \textbf{11}, 297--313.
\newblock (\doi{10.1002/prot.340110408})

\bibitem[{Shih \emph{et~al.}(2007)Shih, Hsiao, Ho \& Li}]{Shih2007}
Shih, A.~C., Hsiao, T.~C., Ho, M.~S. \& Li, W.~H. 2007 Simultaneous amino acid
  substitutions at antigenic sites drive influenza {A} hemagglutinin evolution.
\newblock \emph{Proc. Natl. Acad. Sci. USA}, \textbf{104}, 6283--6288.
\newblock (\doi{10.1073/pnas.0701396104})

\bibitem[{Smith \emph{et~al.}(2004)Smith, Lapedes, de~Jong, Bestebroer,
  Rimmelzwaan, Osterhaus \& Fouchier}]{Smith2004}
Smith, D.~J., Lapedes, A.~S., de~Jong, J.~C., Bestebroer, T.~M., Rimmelzwaan,
  G.~F., Osterhaus, A. D. M.~E. \& Fouchier, R. A.~M. 2004 Mapping the
  antigenic and genetic evolution of influenza virus.
\newblock \emph{Science}, \textbf{305}, 371--376.
\newblock (\doi{10.1126/science.1097211})

\bibitem[{Stewart \emph{et~al.}(1997)Stewart, Lee, Ibrahim, Watts, Shlomchik,
  Weigert \& Litwin}]{Stewart1997}
Stewart, J.~J., Lee, C.~Y., Ibrahim, S., Watts, P., Shlomchik, M., Weigert, M.
  \& Litwin, S. 1997 A shannon entropy analysis of immunoglobulin and {T} cell
  receptor.
\newblock \emph{Mol. Immunol.}, \textbf{34}, 1067--1082.
\newblock (\doi{10.1016/S0161-5890(97)00130-2})

\bibitem[{Valdar(2002)}]{Valdar2002}
Valdar, W. S.~J. 2002 Scoring residue conservation.
\newblock \emph{Proteins}, \textbf{48}, 227--241.
\newblock (\doi{10.1002/prot.10146})

\bibitem[{Wang \& Samudrala(2006)}]{Wang2006}
Wang, K. \& Samudrala, R. 2006 Incorporating background frequency improves
  entropy-based residue conservation measures.
\newblock \emph{BMC Bioinformatics}, \textbf{7}, 385.
\newblock (\doi{10.1186/1471-2105-7-385})

\bibitem[{Wiley \emph{et~al.}(1981)Wiley, Wilson \& Skehel}]{Wiley1981}
Wiley, D.~C., Wilson, I.~A. \& Skehel, J.~J. 1981 Structural identification of
  the antibody-binding sites of {H}ong {K}ong influenza haemagglutinin and
  their involvement in antigenic variation.
\newblock \emph{Nature}, \textbf{289}, 373--378.
\newblock (\doi{10.1038/289373a0})

\bibitem[{Williamson(1995)}]{Williamson1995}
Williamson, R.~M. 1995 Information theory analysis of the relationship between
  primary sequence structure and ligand recognition among a class of
  facilitated transporters.
\newblock \emph{J. Theor. Biol.}, \textbf{174}, 179--188.
\newblock (\doi{10.1006/jtbi.1995.0090})

\bibitem[{Wolf \emph{et~al.}(2010)Wolf, Nikolskaya, Cherry, Viboud, Koonin \&
  Lipman}]{Wolf2010}
Wolf, Y.~I., Nikolskaya, A., Cherry, J.~L., Viboud, C., Koonin, E. \& Lipman,
  D.~J. 2010 Projection of seasonal influenza severity from sequence and
  serological data.
\newblock \emph{PLoS Curr}, \textbf{2}, RRN1200.
\newblock (\doi{10.1371/currents.RRN1200})

\bibitem[{Wu \emph{et~al.}(2010)Wu, Peng, Du, Shu \& Jiang}]{Wu2010}
Wu, A.~P., Peng, Y.~S., Du, X.~J., Shu, Y.~L. \& Jiang, T.~J. 2010 Correlation
  of influenza virus excess mortality with antigenic variation: Application to
  rapid estimation of influenza mortality burden.
\newblock \emph{PLoS Comput Biol}, \textbf{6}, e1000\,882.
\newblock (\doi{10.1371/journal.pcbi.1000882})

\end{thebibliography}

\end{document}


\title{Appendix}
\author{$\mbox{Keyao Pan}^1$ and $\mbox{Michael W. Deem}^{1,2}$\\\\
Department of $^1$Bioengineering and $^2$Physics \& Astronomy, Rice University,\\
6100 Main Street, Houston, TX 77005}

\date{}

\maketitle

\section{Monte Carlo Simulation of the Patterns of Selection and Diversity}
\label{sec:Monte_Carlo}

\begin{figure}
\centering
\vspace{-1in}
\subfigure[]{
\includegraphics[width=2.75in]{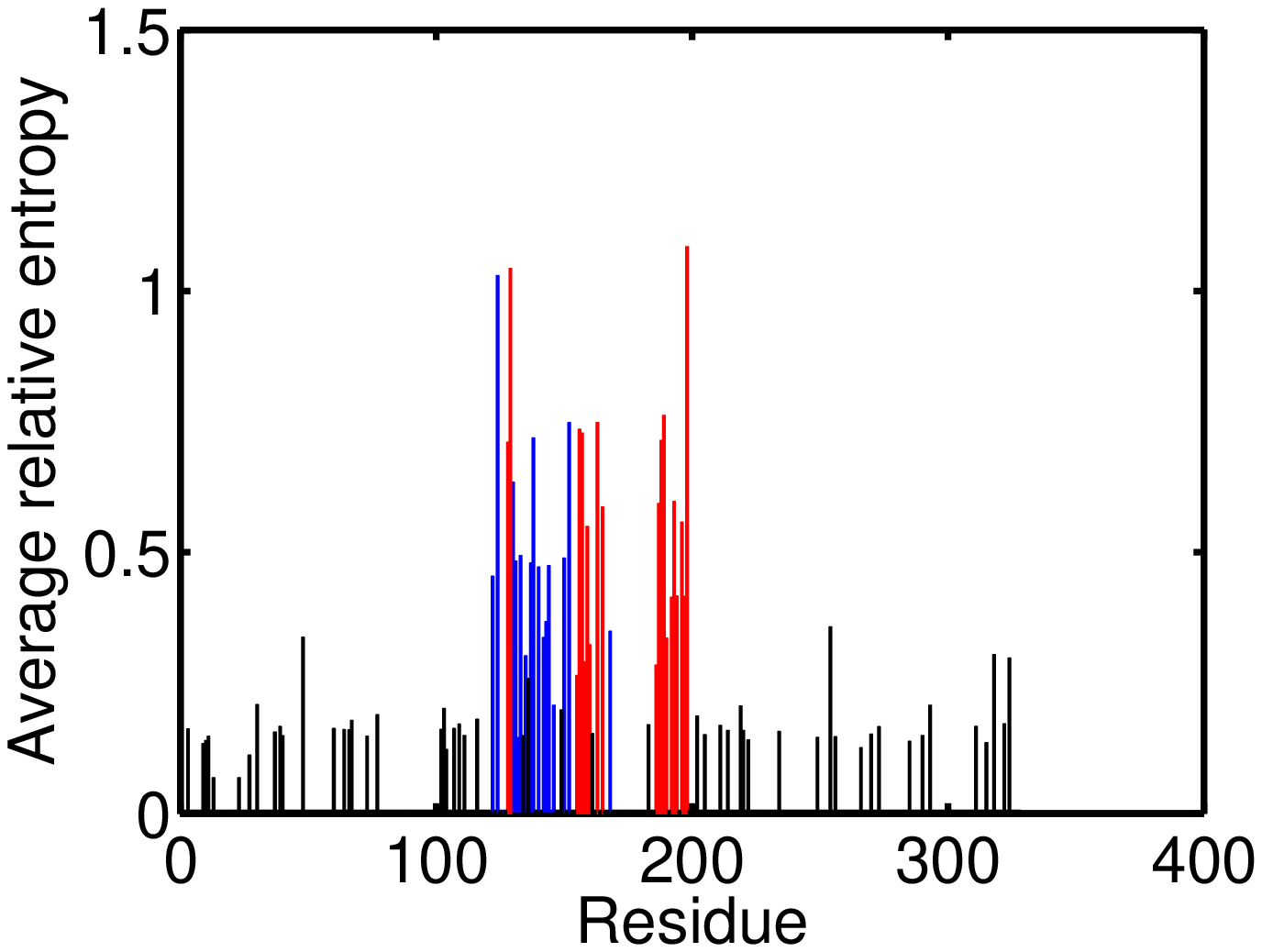}
\label{fig:avg_select_null}
}
\subfigure[]{
\includegraphics[width=2.75in]{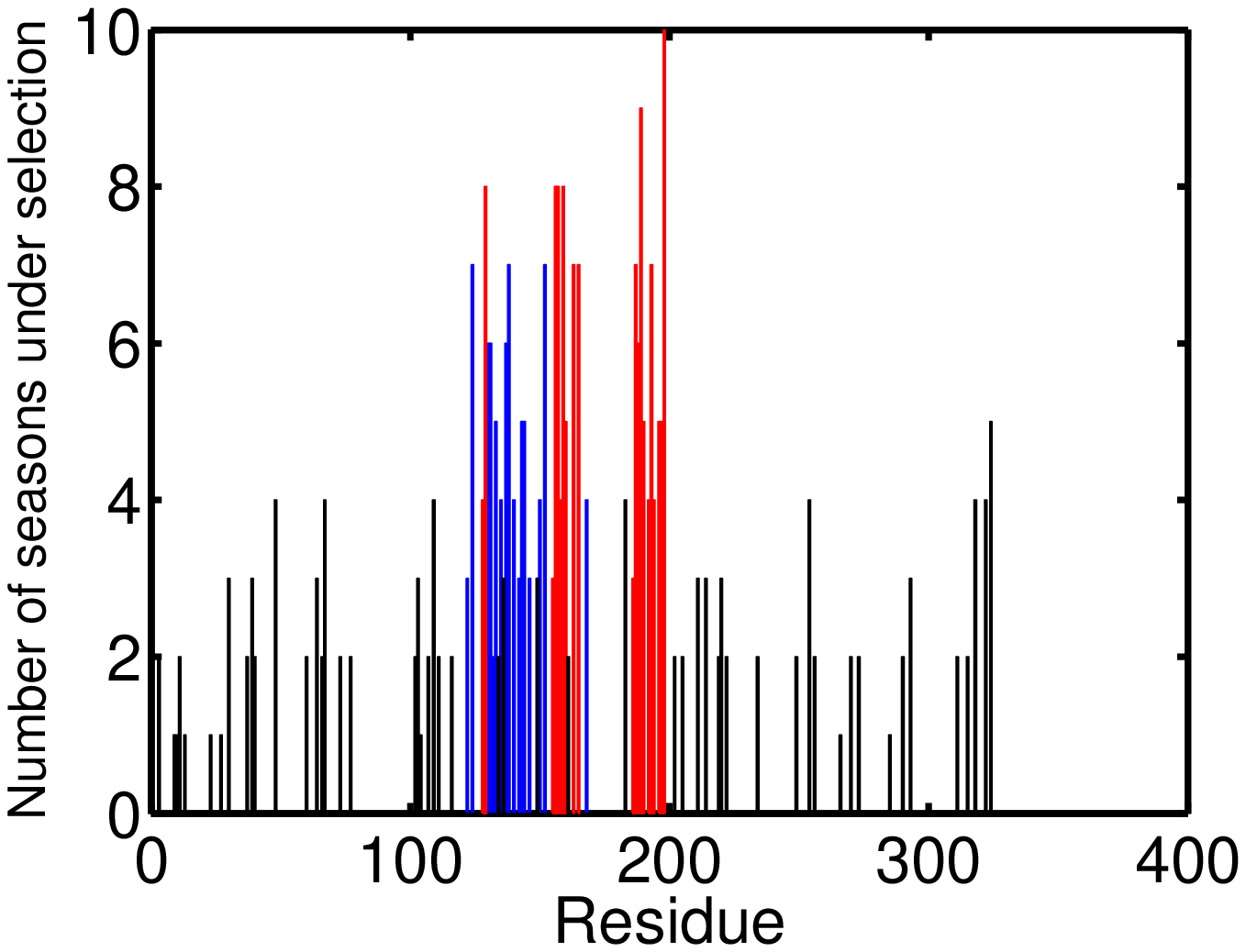}
\label{fig:n_select_null}
}
\subfigure[]{
\includegraphics[width=2.75in]{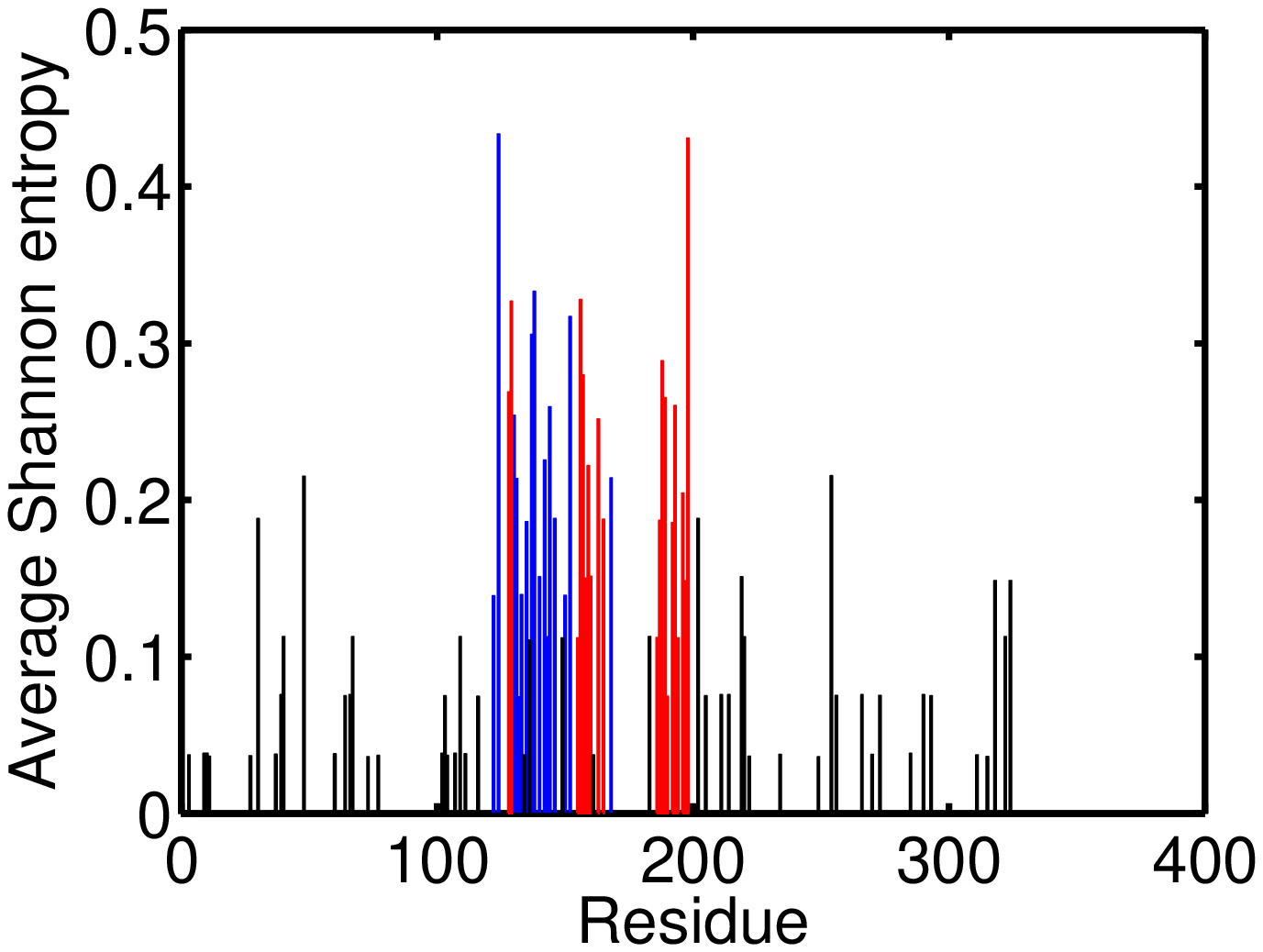}
\label{fig:avg_div_null}
}
\subfigure[]{
\includegraphics[width=2.75in]{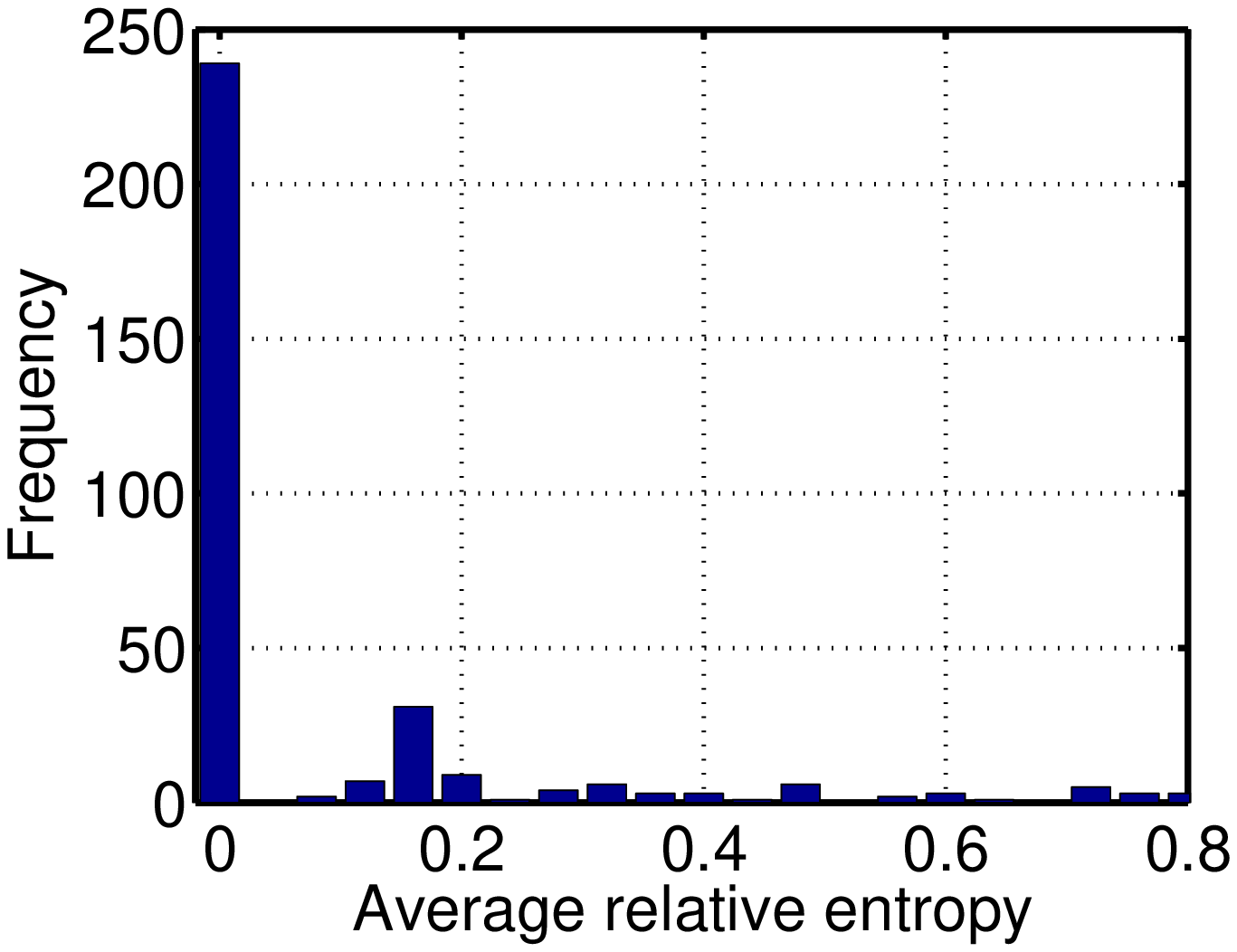}
\label{fig:dist_avg_select_null}
}
\subfigure[]{
\includegraphics[width=2.75in]{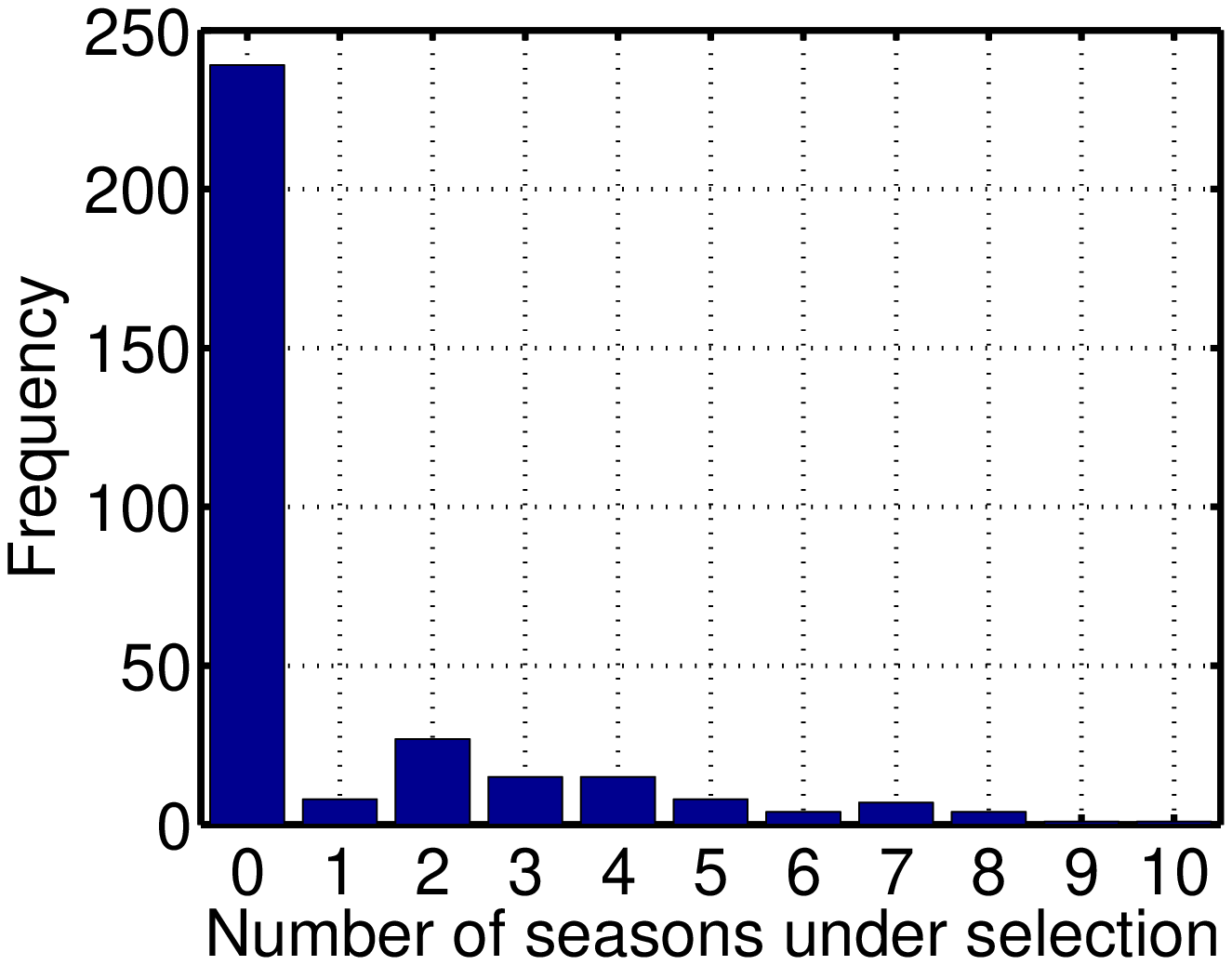}
\label{fig:dist_n_select_null}
}
\subfigure[]{
\includegraphics[width=2.75in]{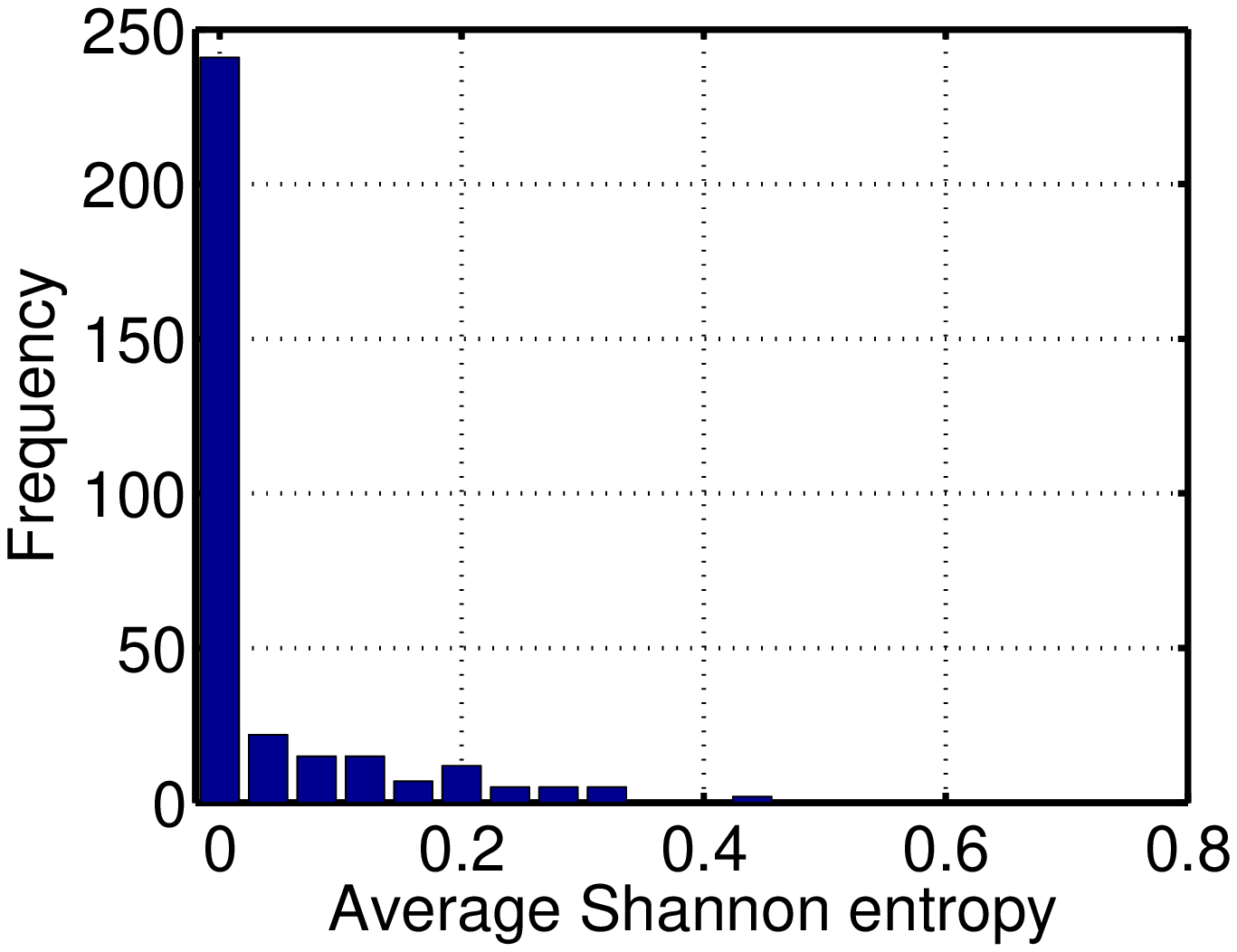}
\label{fig:dist_avg_div_null}
}
\caption{The results of the Monte Carlo simulation model containing epitopes A (blue) and B (red), and all the other positions. The model was simulated for 41 seasons. \subref{fig:avg_select_null} Average selection in each position quantified by relative entropy calculated by $\bar{S}_j = \sum_{i=1}^{17} S_{i,j} / 17$ in the last 17 seasons. The colors represent positions in epitopes A and B and positions outside the epitopes. \subref{fig:n_select_null} Number of seasons for each position when the relative entropy was greater than the threshold $S_i^\mathrm{thres}$, i.e. the position was under selection. \subref{fig:avg_div_null} Average diversity in each position quantified by Shannon entropy in the 17 seasons, calculated by $\bar{D}_j = \sum_{i=1}^{17} D_{i,j} / 17$. \subref{fig:dist_avg_select_null} Distribution of the average selection in each position displayed in \subref{fig:avg_select_null}. \subref{fig:dist_n_select_null} Distribution of the numbers of seasons under selection displayed in \subref{fig:n_select_null}. \subref{fig:dist_avg_div_null} Distribution of the average diversity in each position shown in \subref{fig:avg_div_null}.}
\label{fig:spatial_null}
\end{figure}

We introduce a Monte Carlo model aiming to regenerate the patterns of selection and diversity shown in Figure 4 in the main text.  In this model, the sequence of the HA1 domain contains a dominant epitope bound by the antibody and possessing a high evolutionary rate $\mu_1 = 0.12\ \mbox{amino acid substitution/site/season}$, with the other amino acid positions with a low evolutionary rate $\mu_2 = 0.0034\ \mbox{amino acid substitution/site/season}$ \citep{Ferguson2003}.  In each season, the numbers of positions in and out of the dominant epitope were defined as $L_1$ and $L_2$, respectively.  An ensemble of 1000 HA1 sequences was created with identical amino acid identity in each of the 329 positions, and was simulated from the 1969--1970 season to the 2009--2010 season.  The historical dominant epitopes of H3 hemagglutinin were epitopes A and B, each of which was dominant for about seven seasons \citep{Gupta2006}.  Therefore, the dominant epitope in the model was initialized as epitope A, and shifted between epitopes A and B every seven seasons.  In each season, the numbers of amino acid substitutions in and out of the dominant epitope were randomly determined from two Poisson distributions with mean values $\lambda_1 = \mu_1 L_1$ and $\lambda_2 = \mu_2 L_2$, respectively.  The positions of amino acid substitutions and the new amino acid identities were then randomly assigned.  This process was repeated for all the 1000 sequences in each season.  Following this procedure, we calculated for each position $j$ the average relative entropy $\bar{S}_j$, the number of seasons under selection $N_j$, and the average diversity $\bar{D}_j$, using the simulated sequences between the 1993--1994 season and the 2009--2010 season. We also present the histograms of $\bar{S}_j$, $N_j$, and $\bar{D}_j$.  The results of these calculations are shown in Figure \ref{fig:spatial_null}.

The general picture depicted by this Monte Carlo simulation model reflects natural influenza evolution.  The similarity between the results from the historical sequences in Figure 4 and those from the Monte Carlo simulation model in Figure \ref{fig:spatial_null} suggests that the Monte Carlo simulation model captures a major part of the picture of influenza evolution.  The Monte Carlo simulation model also suggests that the binding of antibody and the increased substitution rate in the dominant epitope are the significant features of influenza evolution.

The annual evolution in the dominant epitope bound by the antibody decreases the affinity between virus and antibody and enables the virus to escape the immune memory of the virus circulating in the previous seasons. The evolutionary rate in the dominant epitope is $\mu_1 = 0.12\ \mbox{amino acid substitution/site/season}$ \citep{Ferguson2003}. Both the historical sequences of the HA1 domain \citep{Shih2007} and the Monte Carlo simulation model suggest that substitutions of amino acid identities have occurred randomly across the dominant epitope. The relative frequency $f\left(k,i,j\right)$ of each amino acid $k$ in each position $j$ in each season $i$ shows that the amino acid substitutions were in random positions in each season and displayed few visible correlation between two positions \citep{Shih2007}. The Monte Carlo simulation model randomly selects the positions in which the amino acid is mutated, and generates similar patterns of selection and diversity to the historical data.

\newpage

\bibliographystyle{rspublicnat}
\bibliography{references}